\documentclass[preprint2]{aastex}
\usepackage{graphicx}
\usepackage{ifthen}

\def\ltsima{$\; \buildrel < \over \sim \;$}
\def\simlt{\lower.5ex\hbox{\ltsima}}
\def\gtsima{$\; \buildrel > \over \sim \;$}
\def\simgt{\lower.5ex\hbox{\gtsima}}
\def\kms{{\rm\,km\,s^{-1}}}

\def\kpc{{\rm\,kpc}}

\def\msun{{\rm\,M_\odot}}

\def\pc{{\rm\,pc}}

\def\CompactFigs{1}
\def\UseFigs{1}

\def\s{\ifmmode \widetilde \else \~\fi}
\def\={\overline}

\def\spose#1{\hbox to 0pt{#1\hss}}

\def\etal{{\it et al.\ }}

\def\lta{\mathrel{\spose{\lower 3pt\hbox{$\mathchar"218$}}
     \raise 2.0pt\hbox{$\mathchar"13C$}}}
\def\gta{\mathrel{\spose{\lower 3pt\hbox{$\mathchar"218$}}
     \raise 2.0pt\hbox{$\mathchar"13E$}}}
\def\Dt{\spose{\raise 1.5ex\hbox{\hskip3pt$\mathchar"201$}}}    
\def\dt{\spose{\raise 1.0ex\hbox{\hskip2pt$\mathchar"201$}}}    

\def\dotsfill{\leaders\hbox to 1em{\hss.\hss}\hfill}

\def\Gyr{{\rm\,Gyr}}

\shorttitle{Ibata et al.}
\shortauthors{Mass distribution of the Halo}

\begin{document}

\title{Great Circle tidal streams:\\
evidence for a nearly spherical massive dark halo around the Milky Way}


\author{
Rodrigo Ibata\altaffilmark{1}, 
Geraint F. Lewis\altaffilmark{2,3,4},
Michael Irwin\altaffilmark{5},
Edward Totten\altaffilmark{6},
Thomas Quinn\altaffilmark{3}}

\altaffiltext{1}{Max-Plank Institut f\"ur Astronomie, 
K\"onigstuhl 17, D--69117 Heidelberg, Germany}
\altaffiltext{2}{ 
Dept. of Physics and Astronomy, University of Victoria, Victoria, B.C., Canada}
\altaffiltext{3}{ 
Astronomy Dept., University of Washington, Seattle WA, U.S.A.}
\altaffiltext{4}{
Anglo-Australian Observatory, PO Box 296, Epping, NSW 1710, Australia}
\altaffiltext{5}{
Institute of Astronomy, Madingley Road, Cambridge, CB3 0HA, U.K.}
\altaffiltext{6}{
Dept. of Physics, Keele University, Keele, Staffordshire, ST5 5BG, U.K.}

\begin{abstract}
An all high-latitude sky survey for  cool carbon giant stars in the Galactic
halo  has   revealed  75  such  stars,   of  which  the   majority  are  new
detections. Of these, more than half  are clustered on a Great Circle on the
sky which intersects the center  of Sagittarius dwarf galaxy and is parallel
to  its proper  motion  vector, while  many  of the  remainder are  outlying
Magellanic  Cloud  carbon  stars.   Previous numerical  experiments  of  the
disruption  of the  Sagittarius dwarf  galaxy (the  closest of  the Galactic
satellite galaxies)  predicted that the  effect of the strong  tides, during
its repeated close encounters with the Milky Way, would be to slowly disrupt
that  galaxy.   Due  to  the  small velocity  dispersion  of  the  disrupted
particles, these  disperse slowly along (approximately) the  orbital path of
the progenitor, eventually giving rise to a very long stream of tidal debris
surrounding  our Galaxy.   The  more recently  disrupted  fragments of  this
stream should contain a similar mix  of stellar populations to that found in
the progenitor, which includes giant carbon stars.

Given the measured position and  velocity of the Sagittarius dwarf, we first
integrate its  orbit assuming  a standard spherical  model for  the Galactic
potential, and find both that the  path of the orbit intersects the position
of the stream, and that the radial velocity of the orbit, as viewed from the
Solar position, agrees very well  with the observed radial velocities of the
carbon  stars. We  also present  a pole-count  analysis of  the  carbon star
distribution, which clearly  indicates that the Great Circle  stream we have
isolated  is statistically  significant, being  a  5-6$\sigma$ over-density.
These two arguments strongly support our conclusion that a large fraction of
the Halo carbon stars originated in the Sagittarius dwarf galaxy.

The stream orbits the Galaxy between the present location of the Sagittarius
dwarf, $16\kpc$ from the Galactic center, and the most distant stream carbon
star, at  $\sim 60\kpc$.  It  follows neither a  polar nor a  Galactic plane
orbit, so  that a  large range in  both Galactic  $R$ and $z$  distances are
probed.  That  the stream is observed  as a Great Circle  indicates that the
Galaxy does not exert a significant  torque upon the stream, so the Galactic
potential  must   be  nearly  spherical   in  the  regions  probed   by  the
stream. Furthermore, the  radial mass distribution of the  Halo must allow a
particle  at the position  and with  the velocity  of the  Sagittarius dwarf
galaxy to reach the distance of  the furthest stream carbon stars. Thus, the
Sagittarius  dwarf  galaxy tidal  stream  gives  a  very powerful  means  to
constrain the mass distribution it resides in, that is, the dark halo.

We  present  N-body experiments  simulating  this  disruption  process as  a
function of  the distribution  of mass in  the Galactic halo.   A likelihood
analysis  shows that, in  the Galactocentric  distance range  $16\kpc <  R <
60\kpc$, the dark halo is most likely almost spherical. We rule out, at high
confidence levels,  the possibility that  the Halo is  significantly oblate,
with  isodensity  contours of  aspect  $q_m <  0.7$.  This  result is  quite
unexpected and contests currently popular galaxy formation models.
\end{abstract}




%

\section{Introduction}

One  of the  most intriguing  puzzles of  modern astronomy  is  the apparent
existence of unseen  mass on galactic and larger scales.  Even for the Milky
Way, our  knowledge of the dark  component is poor, the  little known having
been gleamed from indirect, mostly kinematic, probes.

The  best constraints  on  the mass  distribution  in the  Galaxy have  been
derived from the  kinematics of tracers of the  Galactic disk, in particular
the \ion{H}{1} gas,  which samples the Galaxy from almost  its center to the
outermost  parts of  its disk.   Indeed, the  success of  this  technique in
fitting the  \ion{H}{1} velocities at Galactic longitudes  $\ell > 30^\circ$
strongly  suggests  \citep{dehnen}  that   the  Galaxy  is  axisymmetric  at
Galactocentric distances  greater than $\sim 5  \kpc$ --- where  most of the
mass resides.  Information  on the vertical mass gradient  has been obtained
from the  study of tracers  of the Galactic spheroid  \citep{wyse, binney87,
morrison90,  sommer-larsen, vandermarel,  amendt, morrison00}.   Yet despite
considerable observational and theoretical effort, the vertical structure of
the  spheroid  component  is  still  poorly constrained;  dependent  on  the
technique employed, the  spheroid's oblateness is found to  lie in the range
$0.5 < q  < 1$.  (It is  usually assumed that the density  of the spheroidal
components   of    the   Galaxy   is   of   the    form   $\rho(s)$,   where
$s=(R^2+(z/q)^2)^{1/2}$, and $R$, $z$ are Galactic cylindrical coordinates).
This  paucity of information  on the  vertical structure  of the  Galaxy, it
transpires, is the primary cause  of the large uncertainties on the Galactic
mass models \citep{dehnen}.

The above studies of the stellar spheroid do not, however, necessarily place
strong constraints  on the massive dark  halo.  This is  because the stellar
spheroid  is  a partially  self-gravitating  system,  with  its own  density
profile and kinematics.  Also, at large  radii, the stellar halo will not be
dynamically relaxed, so  we should expect to see  a light distribution which
reflects the formation  process of this component (which  may well have been
quite different to  that of the dark halo).  Strong  constraints on the dark
material   must  therefore   come   from  dynamical   analyses.   At   large
Galactocentric radii,  satellites provide the  only means to probe  the dark
halo \citep{zaritsky89, lin, zaritsky94}. Within  the end of a galaxy's disk
other  analyses are  possible,  probing the  self-gravity  of stellar  disks
\citep{ueda}, or  of HI  disks \citep{becquaert}.  For  the case of  our own
Galaxy, it has  already been possible to place constraints  on the dark halo
distribution through  an analysis of  the HI disk  thickness \citep{olling},
which suggests  that the mass  distribution is not  substantially flattened,
with $q \sim 0.8$.

In this contribution,  we analyze kinematic and distance data  of a group of
newly-discovered stars that  are part of a stream of  material that has been
tidally stripped  from the  Sagittarius dwarf galaxy.   A stream  gives much
stronger  constraints  on  the  underlying  mass distribution  than  does  a
uniformly distributed spheroid star population, as it implies a low velocity
dispersion  and  a  continuity  between  the  stream  members.  It  provides
essentially the trace of an orbit through the Halo.  The stream studied here
is  now distributed  through the  Halo between  Galactocentric  distances of
$16\kpc$ to $\sim 60\kpc$, probing a  region between the end of the Galactic
disk, and the Magellanic clouds.

In section~2,  we discuss  the all-sky survey  that reveals the  Halo carbon
star population; section~3 reviews previous analyses of the evolution of the
Sagittarius  dwarf galaxy and  their predictions  about its  stellar stream;
section~4 makes a  simple first-pass comparison of the  expected stream with
the  carbon  star  dataset  and   discusses  the  probability  of  a  chance
alignment. Section~5  presents a simple  analytic argument to  constrain the
Halo shape  and mass; section~6 provides  a fuller analysis with  the aid of
N-body  simulations  covering  a  range  of  static  Halo  models.  Finally,
section~7 presents the conclusions of this work.

\section{The Halo carbon star sample}

The   APM   Halo   carbon   star  survey   reported   in   \citet{totten98},
\citet{totten00} and  \citet{irwin00} is  now almost complete.   Palomar sky
survey plates  of the whole northern  sky at Galactic  latitudes $|b| \simgt
30^\circ$,  as well  as UK  Schmidt Telescope  plates covering  most  of the
southern sky  at latitudes  $|b| \simgt 30^\circ$  (see Figure~1)  have been
scanned  and analyzed  to search  for giant  carbon stars  (C-stars).  These
intermediate-age asymptotic  giant branch stars  are intrinsically luminous,
rare  ($\approx$1  per  200~sq.~deg)  and  very  red  in  optical  passbands
\citep{totten98},  and  can therefore  easily  be  distinguished from  other
stellar species, making them a useful tracer of recent accretion events.

C-star candidates were selected according  to their magnitudes and colors on
APM scans of UKST and  POSS1 sky survey photographic plates. Candidates were
selected from  color-magnitude diagrams by requiring  that ${\rm B_J  - R >
2.5}$ and ${\rm 11 < R < 17}$, or the equivalent limits in the O and E POSSI
passbands.  The bright limit was  chosen to minimize contamination by nearby
disk, or thick disk carbon stars, since a typical carbon giant with absolute
magnitude  ${\rm M_R  = -3.5}$,  in  general has  to be  in the  approximate
heliocentric distance  range 10--100kpc to  be included in the  survey.  The
faint limit  was chosen because a  complementary deeper survey  to an R-band
magnitude of R=19, for a quasar program covering several thousand sq.~deg of
high latitude sky (Irwin et al.  1991) revealed no carbon stars fainter than
R$\approx$16.   Low  or medium  resolution  spectra  were  obtained for  all
candidate  C-stars,  yielding  spectral  confirmation, and  for  nearly  all
confirmed  C-stars, a  further medium  resolution spectrum  was  obtained to
derive a radial velocity.

An Aitoff equal-area  projection of the positions and  velocities of all the
high  latitude  carbon  star  giants  with  measured  radial  velocities  is
displayed  in  Figure~2; C-stars  located  within  known Galactic  satellite
galaxies  (LMC,  SMC, Fornax,  Carina,  Sculptor)  have  been removed.   The
accuracy of  the velocity data  was estimated from  fitting the peak  of the
cross-correlation function, and was better than $10\kms$ for all stars.

Proper motion measurements for the majority of the Halo C-star sample showed
that none of  the APM-selected carbon stars had  a significant proper motion
\citep{totten00} ruling out the possibility of sample contamination by dwarf
carbon stars.  JHK photometric data have also been obtained for the majority
of the  Halo C-star in an attempt  to derive approximate distances  so as to
correlate the optical  R-band distance estimates with near  IR estimates and
at the same time obtain a  clearer idea of the errors in estimating distance
to isolated C-stars \citep{totten00}.  With the caveat that some C-stars are
enveloped in dusty  shells causing their distances to  be overestimated, JHK
photometry alone results in distance estimates with 2-$\sigma$ errors at the
25\% level.   Figure~5 of \citet{totten00} shows a  good correlation between
R-band and JHK distance estimates  suggesting that for those C-stars without
extant JHK photometry, R-band measures can suffice.

Visual  inspection  of Figure~2  clearly  shows  that  the resulting  C-star
distribution is not at all random  --- either in position or in velocity ---
significant  localized clumps are  seen, with  obvious correlations  in both
distance and radial velocity.

\section{The Sagittarius dwarf galaxy and its tidal stream}

The  Sagittarius  dwarf galaxy  \citep{ibata94},  is  the closest  satellite
galaxy of the  Milky Way; currently only  $25 \pm 1 \kpc$ from  the Sun, and
$\sim 16 \kpc$ from the Galactic center, it is near the point of its closest
approach to the Galaxy.  Such  proximity to the Galactic center induces huge
tidal stresses  on the dwarf,  which must lead to  significant morphological
modification  if   not  its   eventual  destruction.   The   interaction  is
particularly  interesting,  as  it  provides  a  nearby,  and  hence  easily
observable, example of a system of merging galaxies.

Several   numerical  studies   \citep{johnston95,  oh,   piatek,  velazquez,
johnston99} have shown that dwarf satellite galaxies become elongated in the
tidal field  of their  massive companion.  Tidal  debris from  the satellite
remains close to  the plane of the orbit for several  orbital periods if the
host potential is  sufficiently symmetric for the range  of distances probed
by  the satellite orbit.   Since the  Sagittarius dwarf  is on  the Galactic
minor  axis, almost  directly behind  the  Galactic center,  as viewed  from
Earth,  the projected  elongation must  be  aligned with  the proper  motion
vector (and  the projection of the  orbit) to very  good approximation.  The
component of  the proper  motion of the  central regions of  the Sagittarius
dwarf  along  the   direction  of  the  deduced  orbit   has  been  measured
\citep{irwin96,  ibata97}, indicating that  it is  moving northwards  with a
transverse velocity  of $250  \pm 90 \kms$.   A more recent  analysis, using
deep  HST images  close to  the center  of the  dwarf, taken  with  an epoch
difference of 2 years, gives a  slightly higher velocity of $280\pm 20 \kms$
\citep{ibata00a}.

\citet{ibata98a}  performed numerical  experiments to  determine  the likely
structural  evolution  of  the   Sagittarius  dwarf.   Since  `King  models'
\citep{king} fit the structure of present day dwarf spheroidal galaxies well
\citep{irwin95}, they  used these models  to represent the structure  of the
initial pre-encounter  dwarf galaxy. They set  up a grid of  34 King models,
sampling the  plausible parameter  space of concentration,  central velocity
dispersion and central density.  The  evolution of the models was calculated
using an  N-body tree-code algorithm \citep{richardson},  which was modified
to include the  forces due to the assumed fixed Galaxy  potential and due to
dynamical friction, as approximated by the Chandrasekhar formula \citep{BT}.
The  models, comprising  either 4000  or  8000 particles,  were evolved  for
12~Gyr, a  time equal to the age  of the dominant stellar  population in the
Sagittarius  dwarf \citep{fahlman}.   (The  initial mean  position and  mean
velocity of the models was obtained by integrating the orbit of a point-mass
backwards in time for $12 \Gyr$, with dynamical friction reversed).

Figure~3  displays the end-point  structure of  one of  the \citet{ibata98a}
dwarf galaxy models whose final velocity structure matches well the observed
velocity dispersion  and radial velocity  profile of red-giant  stars within
$\sim  30^\circ$ of  the center  of  that galaxy.   In all  of their  N-body
simulations, the modeled  dwarf was found to give rise  to streams of debris
like those  visible in in  Figure~3; these contain  more than $15\%$  of the
initial  galaxy,  equivalent to  a  mass in  the  range  $10^7 \msun  \simlt
M_{debris}  \simlt 2  \times 10^8  \msun$.  So  a substantial  stellar mass,
comparable to the  total mass of the Fornax  dwarf spheroidal galaxy, should
be detectable  as a long  stream of stars  on the sky.  However,  the stream
material will  be mostly very  faint, and its  surface density will  be low,
approximately $15 {\rm \msun/arcmin^2}$.   A practical method to detect such
material is to search for  intrinsically bright stars, such as carbon stars,
that may trace the debris, which would be visible on large-area photographic
surveys.  Unfortunately, C-stars  are quite rare; the Fornax  dSph has $\sim
100$ C-stars \citep{azzopardi}, so the  debris tail of the Sagittarius dwarf
should contain, to  within an order of magnitude, a  similar number of these
stars.

\section{The origin of the C-star stream}

For simplicity,  the simulations of  \citet{ibata98a} assumed a  polar orbit.
However,  it  was   known  even  at  that  time   that  its  orbit  deviates
substantially from  polar.  \citet{lynden-bell}  estimated that the  pole of
its orbit  is located  at $(\ell=94^\circ,b=11^\circ)$, consistent  with the
HST  proper  motion  measurement  \citep{ibata00a}.  Given  the  full  three
dimensional  position  and three  dimensional  velocity  information on  the
center  of the Sagittarius  dwarf, we  investigate its  orbit in  a standard
Galactic  potential.  Figure~4  shows the  projection  of the  orbit of  the
Sagittarius  dwarf galaxy  integrated  in the  spherical Galactic  potential
model of \citet{johnston95} (this was also  the model used by Ibata \& Lewis
1998).  It  is remarkable that  the orbit of  the Sagittarius dwarf  in this
``spherical cow'' Galaxy  model, where we have adjusted  no free parameters,
passes straight through the majority of the observed C-stars.  The projected
distance and radial velocity of the orbital path also give an acceptable fit
to the C-stars.  As the model potential used in Figure~4 is almost spherical
at large  radii, the orbital path  is almost a Great  Circle.  Inspection of
Figure~4 brings out  clearly to the eye the  stream-like distribution of the
C-stars, which seem to follow a Great Circle on the sky.

To  calculate  the  significance  of  this  feature,  we  have  performed  a
pole-count   analysis   \citep{johnston96}   of  the   entire   APM-selected
high-latitude carbon star  sample of 75 objects.  The  results are presented
in Figure 5.  Using the approximate R-band distance estimates for the carbon
stars, the  observed angular coordinates of  the sample were  corrected to a
Galacto-centric  coordinate system, assuming  a solar  distance of  8.5 kpc.
There is  a clear detection of  a highly significant excess  of carbon stars
along a Great Circle with a pole at $90\pm1$, $13\pm1$ (plus the mirror pole
at $270,-13$).  This  is within a few degrees of  the expected orbital track
of the Sagittarius dwarf from:  its measured proper motion; its direction of
elongation;  and also  from  the analysis  of  \citet{lynden-bell} based  on
"Halo"  globular  cluster  systems.   In  addition, there  is  also  another
significant pole  at $174,-3$ (plus mirror  at $354,+3$) that  is a possible
detection of carbon stars associated with the orbit of the Magellanic Clouds
(we defer  discussion of that  feature to a  subsequent paper).  It  is also
worth noting that  the signature of both of  these features is significantly
enhanced  when  making  use  of  the  solar  parallax  correction  described
previously.

For  a completely  random distribution  of carbon  stars we  would  expect a
fraction sin(10$^\circ$) to lie within  10~deg of an arbitrary Great Circle.
Yet  38  of the  75  faint  high latitude  carbon  stars  lie  close to  the
Sagittarius dwarf  track, and 28 out of  the 75 lie close  to the Magellanic
Cloud (MC) track, compared with an expected number of $\approx$13.  However,
this would be a naive statistical assessment for several reasons.  First, if
a  large fraction  of the  carbon stars  do belong  to Sagittarius  (and the
Magellanic Clouds) then the 'background' level estimated above would clearly
be an  overestimate.  Second, the high  latitude sample is  not yet complete
particularly in the  Southern part of the survey  leading to uneven coverage
and  hence non-uniformity of  the 'background'.   Third, the  lower Galactic
latitudes are  almost completely absent  from the carbon  star distribution,
weighting against Great Circle orbital tracks close to the Galactic equator.
Indeed this  latter effect can be clearly  seen in the general  trend of the
contours  with  respect to  Galactic  latitude.   Therefore,  taking a  more
empirical  approach,   the  general   'background'  level  of   the  contour
distribution well away  from the Sagittarius dwarf and  MC poles, but within
60 degrees of  $b = 0$, averages around 10 counts,  with a 1-sigma variation
of approximately  5 counts (we  will return to  the issue of  the background
``noise''  from other  accretion events  in  \S6).  This  suggests that  the
postulated  tidal  debris  for  the  Sagittarius dwarf  and  MC  tracks  are
significant at  approximately 5-6 sigma  and 3-4 sigma  levels respectively,
thereby clearly revealing  for the first time direct  evidence of giant arcs
of stellar tidal debris associated  with Sagittarius dwarf and possibly also
for the MCs.

Thus  it appears  highly  likely that  a  significant fraction  of the  Halo
C-stars, formed  and were once bound  to the Sagittarius  dwarf galaxy.  The
non-uniform  sky coverage of  our survey  does not  affect this  result.  We
performed Monte  Carlo simulations of  a random Halo carbon  star population
(drawn  from   a  spherical  logarithmic   model)  windowed  by   the  field
distribution of plate material shown (for the Southern sky) in Figure~1.  In
1000 random simulations there were no  instances of a peak as significant as
that seen  in Figure~5.   It is, however,  unclear to  us how to  assess the
significance  of  our  putative  Sgr  stream detection  against  a  possible
arbitrarily   non-uniform  Halo   distribution.   Certain   types   of  Halo
sub-structure, such  as the  majority of the  Halo being in  randomized thin
"spaghetti-like" distributions, would have little effect on the significance
of the detected  peak since the effective smoothing  scale around each trial
Great Circle covers  a broad band of some $\pm10$  degrees.  This is clearly
not  the case if  the Halo  carbon star  population is  composed of  a small
number  of large  scale streams,  whose presence  would make  the pole-count
distribution  lumpier  and  so   lower  the  significance  of  the  observed
pole-count  peaks.  So to  avoid accepting  the existence  of the  two large
streams, one has  to postulate the existence of a  background of large scale
streams.  In this  case, decomposing the Halo distribution  into the minimum
number of  streams required to  "fit" the observables  would seem to  be the
best  way to  proceed.  However,  we have  no a-priori  knowledge  about the
maximum  number  of streams  we  may  accept  (indeed, current  cosmological
simulations predict that  hundreds of dwarf galaxy sized  dark matter clumps
inhabit the halos of giant galaxies like the Milky Way, \citealt{moore}), so
in  this situation  it is  not possible  to assess  the significance  of the
observed  pole-count peaks. It  is worth  stressing at  this point  that the
Great Circle Pole count method only  uses two of the phase space constraints
(distance  makes only weak  a contribution).   As we  show in  Figure~4, the
distance   and   radial  velocity   information   corroborate  the   angular
information.  A  better method needs  to be developed  that uses all  of the
available information rather than some of  it; this will be done below in \S6
by comparison to N-body simulations.

\section{Halo constraints: a ``back of the envelope'' approach}

In the  simulations below, the  tidal stream of  the Sagittarius dwarf  is a
very long-lived  feature, remaining confined as a  continuous narrow feature
for much longer than the age of the Universe. The tidally stripped particles
have a low velocity dispersion  (like their progenitor), and once they drift
beyond close  proximity to the  remaining bound clump, the  only significant
gravitational influence they  feel is that of the  Galaxy.  These properties
make  the stream  material  remain close  to the  locus  of the  orbit of  a
point-mass representing the center of mass  of the dwarf galaxy model. It is
clear that this approximation holds best  if the progenitor has low mass (as
suggested by the work of \citealp{gomez-flechoso}).

We may obtain a simple estimate  of the apogalactic distance of the orbit of
the  Sagittarius dwarf,  assuming a  simple spherical  logarithmic potential
model  for the  Halo  $\Phi(r) =  {v_0^2}  \ln(r)$, where  $v_0$ is  the
circular  velocity  of  the  Galaxy.   Taking  a  perigalactic  distance  of
$16\kpc$, and  a present  velocity of  $330\kms$, we find  the roots  of the
equation:
$$u^2+{{2[\Phi(1/u)-E]}\over{L^2}}=0 ,$$
(Equation 3-13 from  \citealt{BT}), where $u=r^{-1}$, $E$ is  the energy per
unit  mass of  the object,  and $L$  is its  angular momentum.   This simple
calculation  yields  an   apocenter  distance  of  $r_{apo}=56\kpc$  (taking
$v_0=200\kms$) and $r_{apo}=42\kpc$ (with  $v_0=220\kms$).  We see from this
that the Sagittarius dwarf galaxy is capable of reaching the distances where
the C-stars are found (see \S3),  and also that the measured velocity of the
dwarf galaxy provides a means to probe the mass of the Halo (parametrized in
this example as a circular velocity).

The discussion in \S4 showed that  there is a strong over-density of C-stars
on  a Great  Circle that  happens  to coincide  with the  projection of  the
expected orbit of the Sagittarius  dwarf in a standard Galaxy potential with
a  spherical halo.  In  a spherical  potential, the  angular momentum  of an
orbiting  test particle  is conserved,  so there  is no  precession;  so all
orbits follow Great  Circles as viewed from the center  of the potential. In
an axisymmetric  (or more complex)  potential, particles on orbits  that are
neither polar or in the equatorial  plane experience a torque which leads to
a precession; viewed from the center of the potential one generally will not
perceive a  Great Circle  path. Therefore, the  fact that the  C-star stream
follows a Great Circle immediately  tells us that the Galactic potential, in
the  regions occupied  by  the stream,  cannot  be far  from spherical.   To
illustrate  this  point we  consider  orbits  in  the flattened  logarithmic
potential   $\Phi(R,z)  =   {v_0^2}   \ln((R^2+z^2/q^2)^{1/2})$.   In   this
potential, an orbit which is close  to being circular precesses at a rate of
approximately \citep{steiman-cameron}:
$$\omega_p(r) = -{{3}\over{2}} \Big({{v_0}\over{r}}\Big) {{(1-q^2)}\over{(1+2q^2)}} \cos(i) ;$$
this approximation is accurate to $\sim  10$\% at $q=0.8$.  The orbit of the
Sagittarius dwarf is not really  circular, so the preceding expression will
give only a rough approximation to  the precession rate.  We take an average
distance  of  $r=30\kpc$  for  this calculation,  and  $i=77.5^\circ$.   The
precession rate for  even a slightly flattened model,  with $q=0.8$, is very
fast, $\approx  60^\circ/\Gyr$; so  even in this  mildly flattened  halo, we
should not expect to see a  long coherent Great Circle stream if that stream
has taken a few $\Gyr$ to disperse.  Clearly, our data have the potential to
strongly constrain the  shape of the dark halo of the  Milky Way. Note that,
unlike the case of polar  ring galaxies \citep{sackett}, the self-gravity of
the stellar stream is negligible.

In the next section we undertake a more quantitative analysis, to find the
Galaxy models allowed by the data.

\section{Halo constraints: N-body simulations}

The previous  analysis clearly demonstrates  that a substantial  fraction of
giant  carbon  stars recently  identified  in  the  Galactic halo  are  both
spatially and kinematically consistent with  them being members of the tidal
stream torn  from the Sagittarius Dwarf  Galaxy. To employ  these tracers in
constraining the mass distribution within  the Galactic halo, however, it is
important to understand the demise of  the Sagittarius Dwarf in a variety of
models  of  the Galactic  potential.   To this  end,  a  suite of  numerical
simulations were undertaken.

\subsection{Galactic potential model}

The  Galactic  mass models  of  \citet{dehnen},  the  most realistic  models
currently  available,  were   used  to  represent  the  Milky   Way  in  our
simulations. These models  contain 5 components: a thin  disk, a thick disk,
an inter-stellar medium, a bulge and a halo, with parameters that are fit to
available observational data.  We find that the path of the stream is almost
entirely insensitive  to the choice  of (plausible) parameters of  the first
four  of these  components, which  is not  surprising since  the Sagittarius
dwarf orbits  at large Galactocentric distance.  For  those four components,
we  choose the  parameters  of  the \citet{dehnen}  model~2,  which gives  a
marginally better  fit to  the their compilation  of available data  than do
their other standard models.  The  stream, however, is sensitive to the Halo
parameters.  The form of the  Halo mass distribution taken by \citet{dehnen}
is:
$$\rho(R,z)=\rho_0 \Big({{s}\over{r_0}}\Big)^{-\gamma}
\Big(1 + {{s}\over{r_0}}\Big)^{\gamma-\beta}
e^{-s^2/r^2_t} , $$
where $r_0$ is  the core (or scale) radius, $r_t$  is the truncation radius,
$\gamma$ is the power-law index in the core, and $\beta$ the power-law index
outside the  core.  Based on the  kinematics of the outer  satellites of the
Milky Way \citep{zaritsky89},  we assume that the Halo  truncation radius is
well outside the apogalactic distance of the orbit of the Sagittarius dwarf.
The parameters $\beta$, $\gamma$, and $r_0$ have to be chosen simultaneously
to give rise to a realistic rotation curve.  We have chosen to work with two
quite  different  families of  Halo  models.  The  first  of  these (H1)  is
motivated by observations of  galaxies with extended HI distributions, these
all  show flat  rotation curves,  implying that  $\beta=2$.  For  these Halo
models we take a small ($r_0=3\kpc$) constant density ($\gamma=0$) core. The
resulting Galaxy rotation curve is  displayed in Figure~6.  Our other family
of  Halo  models  (H2)  is  motivated by  the  cosmological  simulations  of
\cite{navarro}.   Their  ``universal  profile''  is  equivalent  to  setting
$\gamma=1$ and  $\beta=3$. They  also predict the  size of the  scale radius
$r_0$ as a function  of galaxy mass; assuming that the Milky  Way has a mass
interior  to  $200\kpc$  of  $10^{12}\msun$  (see  \citealt{zaritsky99}  and
references therein), then  $r_0 \sim 20\kpc$ (for the  case of the currently
popular $\Lambda$CDM  cosmology). The Galactic rotation curve  of this model
is also displayed in Figure~6.

The remaining two parameters, which are  to be explored, are the the central
density  (which we  rearrange  into the  more  intuitive $v_c(50\kpc)$,  the
circular velocity at $50\kpc$) and the flattening $q_m$.

A computer algorithm  for calculating the potential and  forces due to these
mass models  was kindly provided by  Dr. W. Dehnen, and  was integrated into
our N-body workhorse,  PKDGRAV, a highly efficient and  scalable N-body code
\citep{stadel}.   In all  the following  simulations we  maintained accurate
forces by using an opening angle  of $0.75$ and expanding the cell moments to
hexadecapole order.   Two body relaxation  was suppressed by using  a spline
softening  of 10  pc, such  that  the inter-particle  forces are  completely
Newtonian at 20  pc.  A variable time-step scheme is used  based on the local
acceleration, $\Delta  t < \eta  \sqrt{|\Phi|}/a$, and density, $\Delta  t <
\eta/\sqrt{G   \rho}$,   with  the   accuracy   parameter   $\eta  =   0.03$
\citep{quinn}.  Typically  we used  100000 base steps  to integrate  12 Gyr.
With the  variable time-steps, this  was equivalent to taking  $5\times 10^7$
time-steps.

For each Galactic model, the current center-of-mass of the Sagittarius dwarf
was orbited backwards for 12~Gyr.   At this `initial position', the dwarf is
modeled as a King profile.   This self-gravitating system is then integrated
forwards  for 12~Gyr  within the  fixed Galactic  potential. Our  aim  is to
compare the structure and kinematics  of the resulting stream at the present
day to  the observed C-star  distribution.  Clearly, the final  structure of
the  dwarf galaxy  remnant and  its stream  depends on  the  assumed initial
distribution function, so we next discuss our choice of model.

\subsection{The initial model of the Sagittarius dwarf}

Our  choice  of  a King  model  for  the  initial distribution  function  is
supported  by the  fact  that isolated  Galactic  dwarf spheroidal  galaxies
conform well to this model  \citep{irwin95}. King models can be parametrized
in terms of a total mass, a concentration, and a half-mass radius.

We  first address  the issue  of  the initial  mass $M_{0}$  of the  object.
Previous   numerical   studies   \citep{johnston95,   velazquez,   ibata98a,
gomez-flechoso}  were unable  to  find  a low-mass  model  robust enough  to
survive the Galactic tides for  $\sim 10\Gyr$, and display a final half-mass
radius as  large as  the observed $r_{1/2}=550\pc$  (for instance,  the best
model of  \citealt{gomez-flechoso} has  a half mass  volume a factor  of 100
smaller than the observed half brightness volume). Recently, a new model has
been proposed by \citet{helmi}. Their model~1 gives a good representation of
available observations of  the central of the dwarf  galaxy.  However, their
model predicts  that the tidal  debris stars should outnumber  remnant dwarf
galaxy stars by a factor of 8.   We make a provisional estimate of the total
N-type  carbon star  population in  the central  regions of  the Sagittarius
dwarf   of   $N_{\rm   C-star}   \sim   100$,   derived   by   scaling   the
\citet{whitelock96}  survey   to  the  total   area  $22^\circ\times7^\circ$
\citep{ibata97, cseresnjes}  over which  the dwarf is  now known  to extend.
The  fraction of  mass  disrupted from  the  dwarf galaxy  is then  strongly
constrained by the observed  Halo C-star distribution presented in Figure~2;
we use this argument in a  companion paper \citep{ibata00c} to show that the
\citet{helmi}  study   is  inconsistent   with  the  observed   carbon  star
distribution, as it predicts too much tidal debris.

\citet{zhao} suggested that an encounter with the LMC $2\Gyr$ ago could have
deflected the Sagittarius dwarf from a longer period orbit into the current,
tidally-disruptive, short-period  orbit.  However, the  extended forward and
backward arms  of disrupted C-stars  argues against that possibility  as the
observed structure  is consistent with the  slow loss of  material over many
close  encounters with  the Milky  Way.  Another  alternative,  suggested by
\citet{jiang} is  that the dwarf galaxy  began by being very  massive, up to
$10^{11}\msun$;  orbital  evolution through  dynamical  friction plus  tidal
dissolution  could have  led  to  its present,  much  reduced, state.   This
scenario now  also appears highly unlikely  in the light of  the Halo C-star
distribution presented in Figure~2: if  the present day Sagittarius dwarf is
a mere 1\%  remnant, we would expect  to see some $\sim 3000$  C-stars and a
substantial  population   of  globular  clusters  (since   the  present  day
Sagittarius  also contains  4 globular  clusters) distributed  in  the Halo,
quite unlike what is observed.  It should be noted that this argument may be
weakened if the original dwarf galaxy had a strong gradient in mass to light
ratio, so  that the luminous  material was most centrally  concentrated, and
therefore the last part to be dissolved.

In summary,  we feel that  a more conservative  model with a  modest initial
mass  in the  range  $10^8\msun$ to  $10^9\msun$ \citep{ibata97,  ibata98a},
requiring  less  dark  matter  in   the  dwarf  galaxy,  provides  a  better
representation of our current knowledge of this system.

Our previous N-body work on  this interaction \cite{ibata98a} probed a large
range of the parameter space of  initial King models. We found that only low
concentration models ($c  < 1$) can give rise  to low concentration remnants
(which is  required by the observations).  (Following  \citet{BT}, we define
$c=\log_10(r_t/r_0)$, where $r_t$ is the  tidal radius and $r_0$ is the King
radius of  the system).   For our  dwarf galaxy model  D1 we  have therefore
chosen $c=0.5$ for  the initial King profile.  The  remaining parameter that
must be  chosen is the half-mass  radius. Our earlier work  also showed that
the minor-axis  half-mass radius tends  to decrease as  disruption proceeds,
thus requiring  a large initial half-mass  radius to be  consistent with the
observed  $0.55\kpc$  minor-axis  half-brightness  radius.   The  choice  of
$r_{1/2}=0.5\kpc$,  is a  compromise between  survivability and  final size.
More extended models  tend to disrupt too quickly to leave  a remnant at the
present day.   While these models with initial  $r_{1/2}=0.5\kpc$ do survive
in  the  Galactic  potentials  described  above, they  give  rise  to  final
half-mass radii  that are  a factor  of $\sim 2$  smaller than  the observed
$0.55\kpc$  minor-axis half-brightness  radius  (physically, the  disruption
rate is,  of course, proportional  to the density  -- so the  discrepancy is
worse). The total mass of model  D1 is $10^8 \msun$. Our second dwarf galaxy
model  D2, is  a  copy of  the  \citet{helmi} model~I,  which  give a  final
structure similar  to what is observed,  but at the expense  of a disruption
rate apparently inconsistent with  the observations.  This model has initial
half-mass  radius  $r_{1/2}=1.0\kpc$, and  concentration  $c=0.7$, giving  a
total mass of $5.74\times 10^8 \msun$.

Thus neither of  the models we use for the initial  state of the Sagittarius
dwarf  progenitor  is  entirely  satisfactory.  However,  by  examining  the
differences  between the  simulation results  with  model D1  (which is  too
compact) and model  D2 (which is too susceptible to  disruption) we shall be
better able to understand the actual evolution of the dwarf galaxy.

\subsection{Simulation results}

The dwarf galaxy model is  integrated forward through the Galactic potential
for $12\Gyr$.  The mass of these  dwarf galaxy models is such that dynamical
friction does  not significantly  influence the evolution  of the  orbit, so
this effect  is neglected. In the  majority of models, the  dwarf galaxy was
represented by  $10^4$ particles, although  to ensure that the  results were
not biased  by the  number of particles  employed, several  simulations were
repeated with $10^5$  particles. To compare the simulations  to the observed
C-stars  (which are  stars of  intermediate  age, $T\simlt  6\Gyr$) we  have
selected for  comparison only those  simulation particles that  were located
within a sphere of radius $5\kpc$  from the remnant center at the simulation
time  $6\Gyr$ before  the present.   Any  population that  was disrupted  at
earlier epochs would be too old to be detectable via C-star tracers.

To  summarize, we employed  two different  families of  Halo models,  H1 and
H2. Each  of these  Halo models  was investigated with  three values  of the
circular  velocity ($v_c(50\kpc)=200\kms,220\kms,240\kms$)  and  with eleven
values of the Halo density flattening ($q_m=0.5,0.55,\dots,1.0$). Both dwarf
galaxy models D1 and D2 were  evolved in these 66 Galactic potentials, for a
total of 132 simulations.

Figures~7 to 10 show the end-state  of the dwarf galaxy evolved in Milky Way
potential  models  H1  with   $v_c(50\kpc)=220\kms$;  the  Halo  models  are
progressively  flatter   from  Figure~7  ($q_m=1$),   Figure~8  ($q_m=0.9$),
Figure~9 ($q_m=0.75$), to  Figure~10 ($q_m=0.5$). Expectations regarding the
effect of the Halo flattening on  the observed structure of the stream (\S5)
are clearly confirmed.  A visual  comparison of these diagrams with the data
(Figure~2)  also appears to  bear out  our earlier  deduction that  the Halo
cannot be  substantially flattened: Figures~7 and  8 seem to  be much better
approximations to the data than Figures~9 and 10.

\subsection{Quantitative comparison to data}

A  straightforward way  to compare  our simulation  results to  the observed
C-star distribution is to compute the likelihood of the various models given
these data.  We  apply Gaussian smoothing to each  simulation particle, using
the  observational  uncertainties in  velocity  ($\Delta  v  = 10\kms$)  and
distance  (${\rm \Delta  R =  0.3  mag}$), and  we further  smooth in  right
ascension and declination with a  dispersion equal to approximately half the
width of the  simulated stream ($\Delta \alpha =  5^\circ$, $\Delta \delta =
5^\circ$); this  gives a better approximation to  the underlying probability
distribution  function  $l_m(\alpha,\delta,{\rm R},v)$  of  model $m$.   The
likelihood  of model  $m$ is  then  the product  of the  likelihoods of  the
(four-dimensional) data points:
$$L_m({\cal D}) = \prod_{i=1}^{n} l_m(d_i) \, ,$$
where  $d_i$ is  one of  the $n$  elements of  the data  set  ${\cal D}_{\rm
real}=\{d_1,\dots,d_n\}$.  This technique provides a relative ranking of the
models, and  allows us to  select the best  model, $M$ say,  which maximizes
$L$.  However, to  have a meaningful result, we also  need to ascertain that
our  best model  provides  an  acceptable representation  of  the data.   To
investigate this,  we use the  probability distribution function  $l_M$ (the
smoothed  simulation end-state  of  model $M$)  to  generate many  synthetic
realizations  of our  data, ${\cal  D}_{\rm M}=\{d_1,\dots,d_n\}$  (the same
selection criteria are,  of course, applied to the synthetic  data as to the
real data).  Comparing  $L_M(\cal D)$ calculated from the  observed data, to
the distribution  of $L_M(\cal D)_{\rm  M}$ derived from the  synthetic data
sets, gives the probability that a likelihood as low as $L_M(\cal D)$ should
occur by chance if model $M$ were correct.

We applied this procedure to  the simulations discussed above and found that
all  models can  be rejected  at very  high confidence  level (worse  than 1
chance  per million).  However,  in the  discussion above,  we did  not take
account of outlying  data points.  It is clear that the  presence of data in
phase-space regions  uninhabited by  the model will  lead to very  low model
likelihoods, such as would never occur  by chance if the model were correct.
As discussed above, these outliers  may trace other recent accretion events.
In principle, a background model  for these extra events can be incorporated
into our probability  distribution function, but we refrain  from doing this
as  we have  no  good model  for  the four-dimensional  ($\alpha,\delta,{\rm
R},v$) distribution of these ``contaminants''.

Instead, we  attempt to  reject the contaminants  as follows.   We calculate
$l_m(d_i)$  for the  data set  ${\cal D}_{\rm  real}=\{d_1,\dots,d_n\}$, and
sort these data  points into order of descending  likelihood. This allows us
to     construct     subsets      of     the     data     ${\cal     D}_{\rm
real}^{-k}=\{d_1,\dots,d_{n-k}\}$, for  which the $k$ worst  data points are
ignored (that is, they are  considered to be background events).  As before,
we use the  model to construct many synthetic realizations  of the data, but
now the synthetic  data sets contain only $n-k$  data points.  Starting with
$k=0$, we successively reduce the number  of data points $n-k$ until we find
a model which cannot be rejected at the 90\% confidence level.

Applying this procedure, we find that  the maximum number of points that can
be picked from  the full survey set of $n=60$  C-stars (with radial velocity
measurements), and be consistent with  any of our models is $n-k=19$~.  This
would suggest that at least $\approx 1/3$ of the Halo C-stars are associated
to  the  Sagittarius  dwarf  tidal   stream.   The  most  likely  model,  it
transpires, is the D2 dwarf galaxy model simulated in the H2 Milky Way model
with $v_c(50\kpc)=200\kms$  and $q_m=1.0$.   Models with a  nearly spherical
Halo  are strongly  preferred by  the data  over oblate  models ---  this is
clearly seen  from inspection of  the likelihood curves of  Figure~11.  Halo
flattening values more oblate than $q_m  < 0.7$ can be rejected at very high
confidence levels (less than 1 chance per million).

Due to  the expense of  running the N-body  simulations, the grid  of Galaxy
mass  models we  employ  here probes  only  3 values  of $v_c(50\kpc)$,  the
circular velocity at $50\kpc$. Of  these, the data clearly indicate that the
lowest  value of  $v_c(50\kpc)=200\kms$, is  to be  preferred.   However, we
defer a full analysis of the constraints of the carbon star dataset on $v_c$
to a future contribution, where we  will present a suite of simulations that
better resolves this parameter.

Of course, the  galaxy interaction we have simulated  here has been modelled
in  a  highly idealized  fashion.   The  modelled  Galaxy provides  only  an
unevolving fixed  potential, and  it does  not react to  the passage  of the
Sagittarius dwarf. In reality the relevant components, the Galactic halo and
disk,  will   clump,  and  in   the  case  of   the  disk,  may   even  warp
\citep{ibata98b} in the  presence of the companion, as  they presumably also
do, but less readily, in isolation. The presence of other massive members of
the  Galactic satellite  system,  notably the  Magellanic  Clouds, has  been
ignored in our simulations.  The dwarf galaxy model is also very simple: its
dark matter profile  is taken to be  a King model, and we  have not included
the accretion of  disk (or other) gas onto it  \citep{ibata98b}, or for that
matter,  the  ejection of  gas  during  disk  crossings.  However,  all  the
simplifications we  have taken  lead to a  smoother, less  eventful modelled
interaction,  with fewer  variable accelerations  than actually  happened in
reality. The  inclusion of the  above complications into  future simulations
will introduce stochastic non-radial  forces, which will necessarily cause a
faster precession  of the stream and,  for a given Halo  flattening, cause a
greater deviation from the observed great circle stream.  For this reason we
consider that our analysis places a lower limit on the Halo's shape.

\subsection{Verification of the statistical method}

The  above method,  which  is effectively  a  sigma-clipping algorithm,  was
implemented to  avoid outlying  data points, presumably  ``background'' Halo
stars,  from dominating  the statistics.  Alternative approaches  could have
been taken:  for instance, we  could have smoothed the  simulation particles
with a non-Gaussian kernel, or  included a model for the background sources.
The procedure we adopted above, however, is better in the sense that we have
had to provide no extra ad-hoc parameters (like a choice of smoothing kernel
or a model of the background).

However,  it is  necessary  to  check whether  any  statistical biasses  are
introduced by the method. To accomplish this check, we perform the following
Monte-Carlo  tests   (Test~1,  Test~2,   Test~3),  where  we   simulate  the
observation of the carbon star  sample from some simple Galaxy models.  {\sl
All of  the artificial samples are chosen  to have the same  sky coverage as
the real sample, as well as the same magnitude (distance) limits}.

\subsubsection{Test~1: real sample used for background}
The  first test  we  perform is  to use  the  $k=41$ carbon  stars that  are
rejected by the method of \S6.4  for a background sample.  We draw $n_{Sgr}$
particles from the  end point of the D1 N-body  integrations to simulate the
Sagittarius  stream stars, and  complement these  with randomly  drawn stars
from the background sample to obtain  a total number of 60 objects (the same
number as in  the real dataset).  To avoid multiple  entries of stars during
the random selection of the  background sample, on the second and subsequent
occasions a  given star  is selected, we  simply rotate the  star's position
azimuthally about  the Galactic minor axis  by a random angle  between 0 and
$2\pi$.  The same  analysis is then performed on  1000 datasets simulated in
this way as  on the real data, selecting, as before,  the best fitting model
from  the full  range  of  H1 halo  models  considered above  ($v_c(50\kpc)=
200\kms, 220\kms, 240\kms$; $q_m=0.5, 0.55, \dots, 1.0$).  We performed this
analysis for  five different stream  to background ratios  ($n_{Sgr}=20, 30,
40, 50,  60$).  The resulting halo  flattening error as a  function of input
halo flattening  and input circular  velocity is displayed in  Figure~12 for
the worst  case considered  ($n_{Sgr}=20$): the typical  error is  less than
$\delta q = 0.06$.

\subsubsection{Test~2: Halo model for background}
For our  second test, we chose  a self-consistent approach, in  which a Halo
model is used to provide both  a fixed potential for the dwarf galaxy N-body
simulations, and a description of the  background in our sample. In order to
draw background particles with velocities as well as positions, we of course
need to  know the distribution  function (DF) of  the model halo.   For this
reason we  chose the  lowered Evans' model  (LEM) to represent  the Galactic
halo \citep{kuijken},  a model  which has  a tractable DF  and is  finite in
extent.   We impose  a  tidal radius  of  $200\kpc$, given  that  this is  a
reasonable   estimate   of  the   minimum   size   of   the  Galactic   halo
\citep{zaritsky99}, and  choose a core radius  of $3\kpc$ (to  be similar to
the Halo  model H1 in  \S6.1). The dwarf  galaxy model D1  comprising $10^4$
particles was evolved in 33 different LEM Halo potentials, covering the grid
of  circular  velocity  $200\kms,   220\kms,  240\kms$,  and  {\sl  density}
flattening  $q_m=0.5,  0.55,  \dots,  1.0$  (here,  we  define  the  density
flattening  $q_m$  to be  the  axis ratio  of  the  isodensity surface  that
intersects  the  Galactic plane  at  $R=50\kpc$).   Artificial datasets  are
constructed by drawing $n_{Sgr}$ particles from the N-body simulations, with
a  further  $60-n_{Sgr}$  particles   sampled  from  the  halo  model.   The
background halo was sampled using the algorithm of \citet{kuijken}.  We find
that for $n_{Sgr} \ge 20$, the typical flattening error is less than $\delta
q = 0.08$.

\subsubsection{Test~3: Halo stream model for background}
A  third  test   is  performed  to  investigate  the   effect  of  a  clumpy
background. We  use the  same Halo potentials  and N-body simulations  as in
Test~2, to select as before,  $n_{Sgr}$ stream particles. However, this time
we  only select  three phase-space  points from  the LEM  DF.   These points
define the starting positions and orbits of three hypothetical Halo streams.
To populate the streams, we first select randomly which of the three streams
a particle belongs  to.  We then choose  a time at random between  0 and the
orbital  period  of the  stream,  and place  the  particle  at the  position
corresponding to that time elapsed from the starting position on the guiding
center orbit.  Thus each stream  is uniformly populated along its orbit, and
the  three  streams  have  a  similar  numbers  of  particles.  A  total  of
$60-n_{Sgr}$ particles are selected in  this way to populate the three model
streams.   Figure~13  shows the  result  when  $n_{Sgr}=20$; the  flattening
errors are again acceptable, with $\delta q < 0.1$.

These three  tests show that the  statistical analysis in \S6.4  on the real
carbon  star data  does not  introduce  significant biases  to the  derived
flattening $q_m$.  In particular, the  analysis is able to recover the input
Halo flattening,  even if the carbon  stars from the  Sagittarius stream are
outnumbered 2 to  1 by the Halo background. The background  may be smooth or
clumpy (as would  be expected if there were a small  number of other streams
just  below our present  detection limit).   Neither is  our sky  coverage a
significant  problem.   The tests  therefore  support  the  result that  the
Galactic halo cannot be substantially flattened.

\section{Conclusions}

The results  presented above  provide strong evidence  that the  dark matter
halo   surrounding  our   Galaxy   is  not   significantly  oblate   between
Galactocentric radii of $\sim 16\kpc$  to $\sim 60\kpc$.  Flat halos $q<0.7$
are  ruled  out at  very  high  confidence  levels. Therefore,  dark  matter
candidates  such as  cold molecular  gas  \citep{pfenniger94a, pfenniger94b,
combes} or  massive decaying neutrinos \citep{sciama}, that  would give rise
to  a highly  flattened component,  cannot contribute  significantly  to the
galactic mass budget.  Our result is at odds with  some earlier studies that
used spheroid  tracers to  probe the  dark halo; the  large spread  of $q_m$
between different  teams adopting  that approach is  likely a result  of the
different  assumptions  employed to  reduce  the  complexity  of the  Jeans'
equations, and also  of the technical difficulty of  measuring the necessary
input quantities to those equations.  In contrast, our measurement hinges on
the very simple physical principle  of conservation of angular momentum in a
spherical potential, which, we believe, provides a much cleaner test.

\citet{olling}  have recently  summarized  extant measurements  of the  dark
matter shapes of galaxies. The methods used to date are confined to analyses
of the  flaring of the galactic gas  layer, warping of the  gas layer, X-ray
isophotes, polar  ring analysis, and  precession of dusty disks.   They note
that the derived answer correlates strongly with the technique employed (see
their Figure~1). It is interesting  that these measurements are derived from
data at  smaller galactocentric distance  than the stellar  stream presented
here. It  is possible that this reflects  a radial gradient in  the shape of
dark  matter  halos,  the   inner  regions  being  more  flattened.   Better
statistics,   perhaps   from    gravitational   weak   lensing   experiments
\citep{fischer}, may help solve this issue.

Is our result unexpected on theoretical grounds?  Numerical simulations with
purely  collisionless matter  \citep{katz91a,  dubinski91, warren,  katz91b,
summers, dubinski94}  produce flattened triaxial halos,  with a distribution
of flattening ratios that peaks near  $q_m = 0.7$. However, it is found that
the presence of  a realistic fraction of dissipational  gas particles in the
simulations  \citep{katz91b, summers,  steinmetz} alters  the orbits  of the
dark matter  particles, giving  rise to oblate  and slightly  flatter halos.
The conclusion  of this study, albeit  a single measurement,  is somewhat at
odds with these widely popular theories;  we find in all possible cases that
the Galactic  dark matter  halo is closer  to being spherical.   The initial
conditions,  and  the  physics  and  numerical techniques  employed  in  the
subsequent  evolution  of  those  galaxy  formation  simulations  should  be
re-addressed in the  light of these observational results  if other galaxies
are also deduced to have nearly spherical mass distributions at large radii.

Further  surveys should attempt  to discover  the other  stellar populations
associated with the Great Circle  streams identified in this study. Stars of
interest will  include RRLyrae variables, which are  better standard candles
than  C-stars, and  being  much older,  trace  the more  ancient history  of
disruption events in our Galactic halo. These stars could be identified from
the Sloan  Digital Sky  Survey (SDSS) for  instance.  Furthermore,  SDSS (or
other surveys) could help improve  the stream statistics by finding the much
more numerous fainter stars  spatially and kinematically associated with the
C-stars presented above.
\footnote{After   the  submission   of  this   article,   \citet{yanny}  and
\citet{ivezic} discovered  the presence of  a large population  of A-colored
Halo stars in the SDSS dataset,  distributed apparently in a ring around the
Milky Way. In a companion  paper \citep{ibata00b}, we argue that those stars
are most likely the RR-Lyrae  members of the Sagittarius stream discussed in
the  present contribution.  Analysis of  the  (as yet  very limited)  region
covered by the  SDSS survey shows a distribution in  good agreement with the
simulation of Figure~8 (where $q_m=0.9$).}

On a  large telescope equipped  with a wide  field camera (such  as Subaru's
Suprime-Cam), it  will also  be possible to  conduct a  complementary C-star
survey around our  neighboring galaxy M31.  It will  be very interesting to
see  whether  similar  Great Circle  streams  are detected  in  a  different
environment.

With the advent  of the next generation of  space astrometric missions, such
as the National Aeronautics and  Space Administration's SIM and the European
Space Agency's GAIA,  it will be possible to  obtain accurate proper motions
and  even distances  for  stars from  the  Halo streams  identified in  this
contribution.    With   the  resulting   full   6-dimensional  phase   space
information,  we can  expect to  be able  to map  the distant  Galactic mass
distribution  in exquisite detail,  and clarify  how the  Milky Way  and its
satellite galaxies  formed, and  how the  latter came to  be torn  apart and
their contents flung across the sky.

\section*{Acknowledgements}

GFL acknowledges  partial support from  the Theodore Dunham  Research Grants
for Astronomy.

\onecolumn

\begin{figure}
\ifthenelse{\UseFigs=1}{
\ifthenelse{\CompactFigs=0}
{\includegraphics[width=13cm,angle=270,bb= 60 40 580 700,clip]{Halo.fig01.ps}}
{\includegraphics[width=13cm,angle=270,bb= 60 40 580 700,clip]{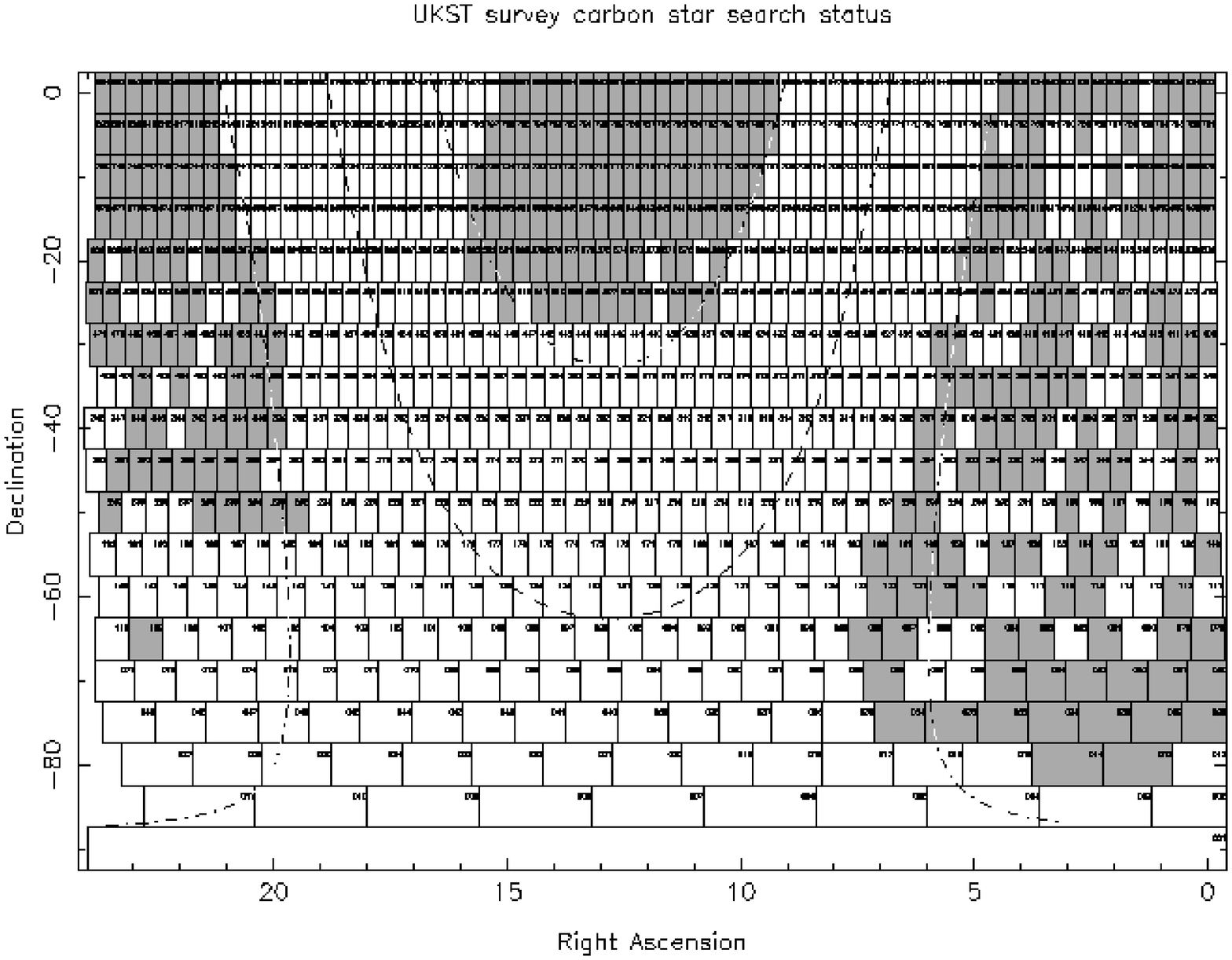}}}{}
\figcaption{The southern UK  Schmidt Telescope photographic regions surveyed
are indicated in  grey. The dashed line marks the  Galactic plane, while the
dot-dashed lines mark $b=30^\circ$ and $b=-30^\circ$.}
\end{figure}

\begin{figure}
\ifthenelse{\UseFigs=1}{
\ifthenelse{\CompactFigs=0}
{\includegraphics[width=13cm]{Halo.fig02.ps}}
{\includegraphics[width=13cm]{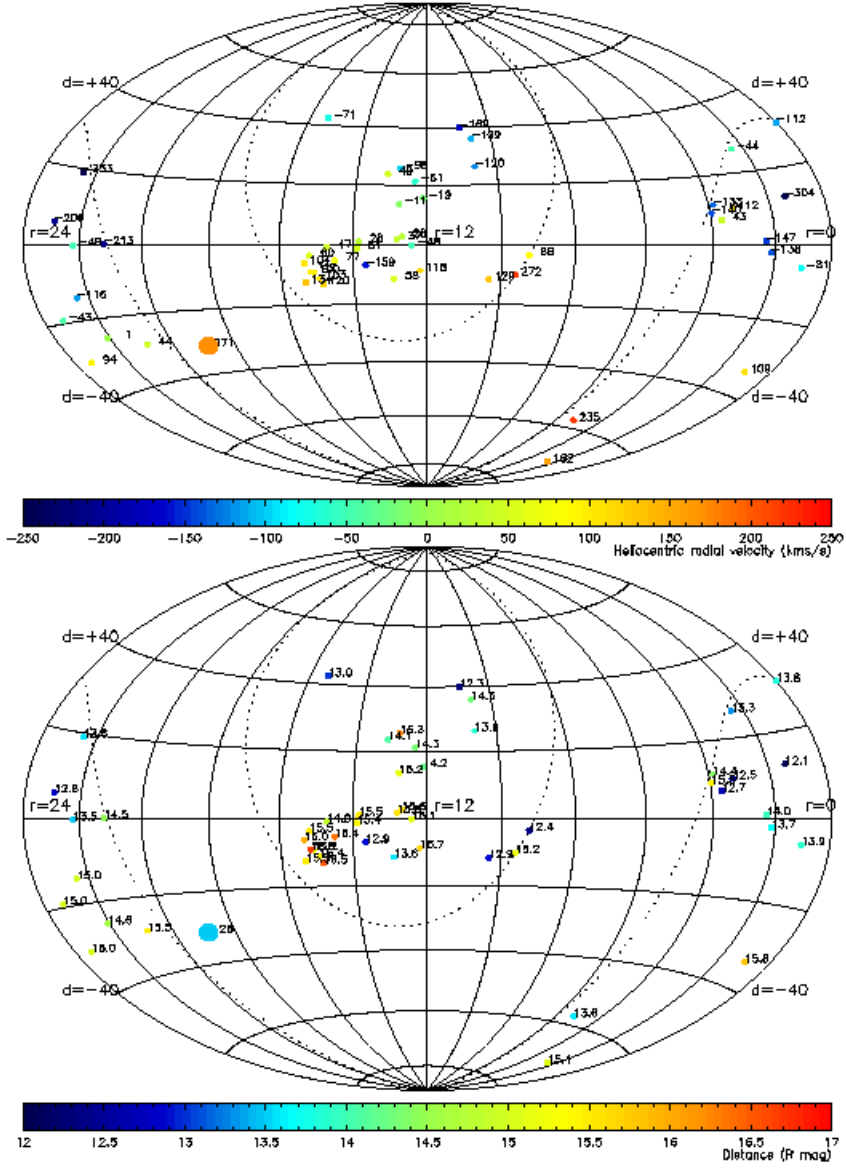}}}{}
\figcaption{Aitoff  projections of  the positions  of the  $60$  Halo carbon
stars  in the  sample with  radial velocity  measurements.  The  upper panel
displays color-coded Heliocentric radial  velocities, while the bottom panel
shows  the stellar  distances, color-coded  as  a function  of the  observed
apparent magnitude.  The apparent  magnitudes are true R-band magnitudes for
those  stars for which  only R-band  data is  available, otherwise  they are
JHK-derived distances from  \citet{totten00}, converted to R-band magnitudes
assuming  ${\rm M_R=-3.5}$.   Given a  representative absolute  magnitude of
${\rm M_R =  -3.5}$ for these objects, the magnitude  range displayed in the
color wedge ranges from $12.5\kpc$  to $125\kpc$.  The large point indicates
the position  and velocity of the  central regions of  the Sagittarius dwarf
galaxy.   Its heliocentric distance  ($25\kpc$) is  labelled on  the diagram
(all other points  are labeled by the R-band  magnitude of the corresponding
C-star).}
\end{figure}

\begin{figure}
\ifthenelse{\UseFigs=1}{
\ifthenelse{\CompactFigs=0}
{\includegraphics[width=\hsize]{Halo.fig03.ps}}
{\includegraphics[width=\hsize]{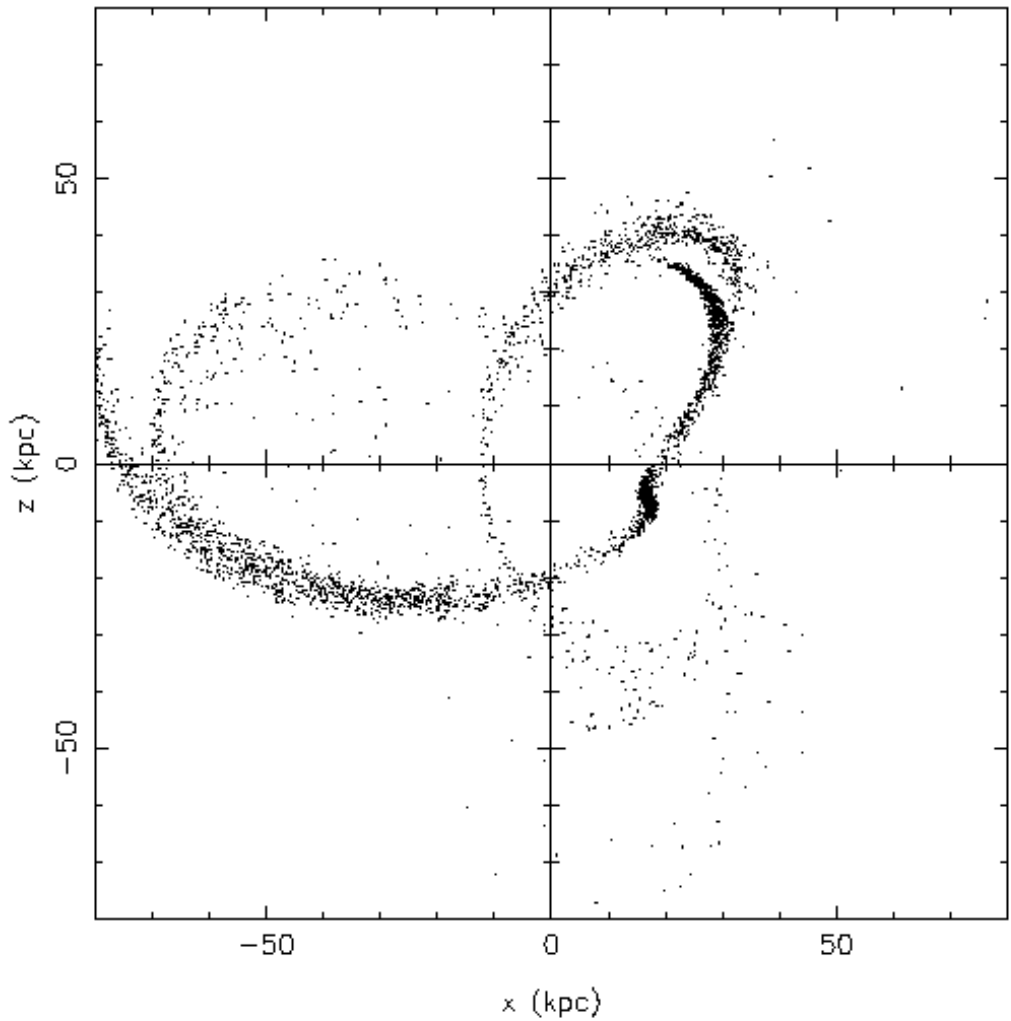}}}{}
\figcaption{The   remnant  of   one  of   the  dwarf   galaxy   models  from
\citet{ibata98a}  (initial  central  density  $\rho_0  =  2.0  \msun/\pc^3$,
initial concentration $c = 0.7$, initial central velocity dispersion $\sigma
= 40 \kms$)  is shown in the  $x$-$z$ plane of the Milky  Way.  The Galactic
center lies at the origin, and the position  of the Sun is at $x = -8 \kpc$.
The dwarf galaxy, evolved for $12  \Gyr$ in the Milky Way potential model of
\citet{johnston95}   (a  spherical   halo   has  been   used),  has   become
significantly disrupted by the Galactic tides; long streams of material lead
and trail  the dwarf  galaxy, which now  contains $<  60 \%$ of  the initial
mass.}
\end{figure}

\begin{figure}
\ifthenelse{\UseFigs=1}{
\ifthenelse{\CompactFigs=0}
{\includegraphics[width=13cm]{Halo.fig04.ps}}
{\includegraphics[width=13cm]{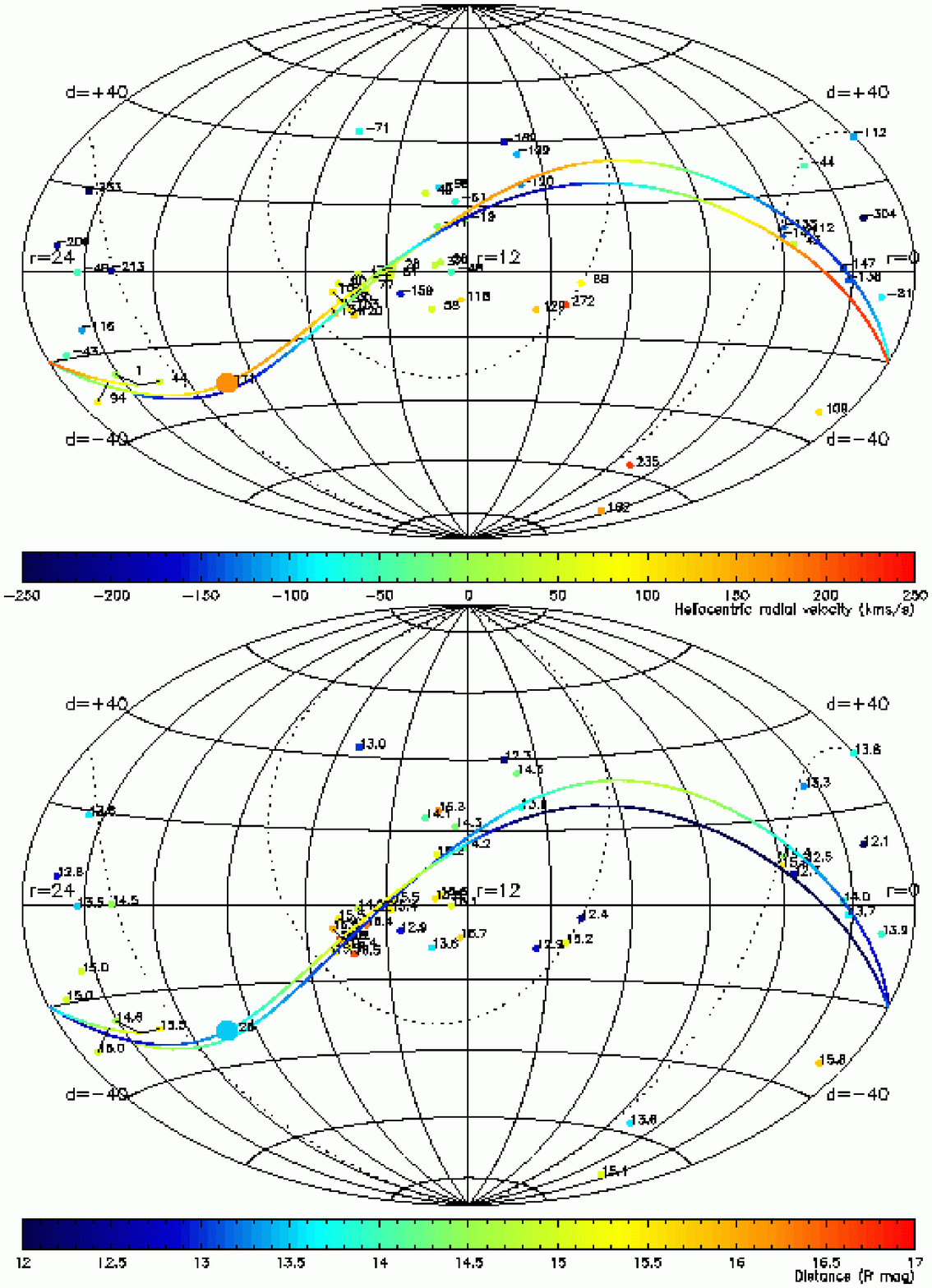}}}{}
\figcaption{Projected  orbit  color-coded  with  velocities  and  distances
overlaid on the data of Figure~2. The orbit has been integrated forwards and
backwards in time for $1\Gyr$ (slightly more than two perigalactic passages)
in  the Galactic  potential used  by \citet{johnston95}  and \cite{ibata98a},
which has a  spherical halo component.  The precession of  the orbit in this
model  is due  to  the disk  component,  which was  modelled  as a  Miyamoto
disk.   At  large   Galactocentric  distance,   this  disk   model  provides
unrealistically  strong  vertical  potential  gradients  (the  disk  extends
indefinitely). With a more realistic  disk potential (as is used below) much
less precession is seen in a Galactic potential with a spherical halo.}
\end{figure}

\begin{figure}
\ifthenelse{\UseFigs=1}{
\includegraphics[width=14cm,angle=270,bb= 60 90 580 720,clip]{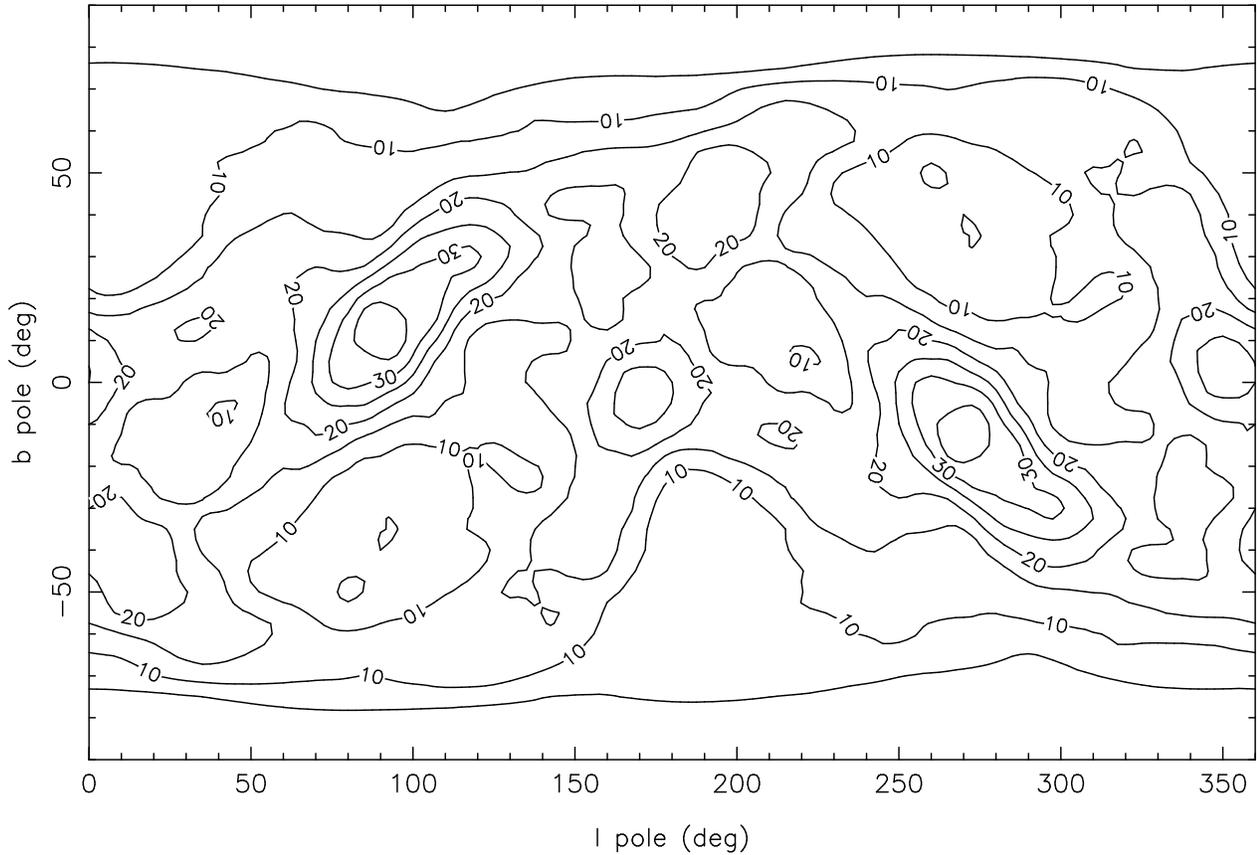}}{}
\figcaption{The contours represent  the number of carbon stars  from the APM
survey, found within 10 degrees of putative Great Circle orbital paths. Each
Great Circle  track is defined by  the Galactic coordinates of  its pole.  A
2.5 degree grid of these possible orbital  poles was used as a basis for the
contour map.  The obvious concentration of  carbon stars at $l = 90^\circ$ ,
$b  =  13^\circ$  is  coincident  with  the expected  orbital  pole  of  the
Sagittarius  dwarf galaxy, while  that at  $l=170^\circ$, $b=-3^\circ$  is a
possible  detection  of stellar  debris  from  the  Magellanic Clouds.   The
observed $l,b$ of  the carbon stars have been  corrected to a Galactocentric
coordinate system  using approximate  distances derived from  their apparent
R-band magnitudes.}
\end{figure}

\begin{figure}
\ifthenelse{\UseFigs=1}{
\includegraphics[width=\hsize]{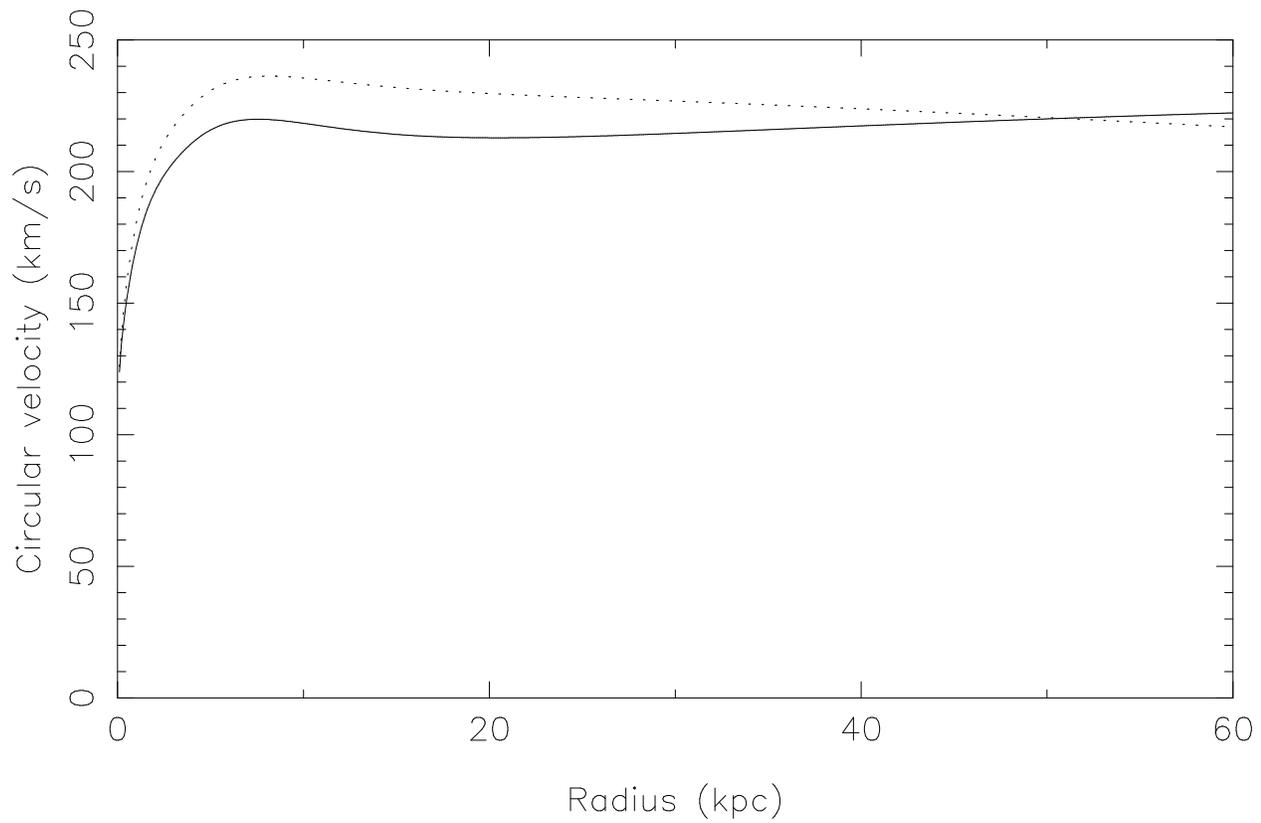}}{}
\figcaption{Total Galactic  rotation curves for  the H1 (solid line)  and H2
(dotted line)  Halo mass  models. In  this diagram, the  Halo mass  has been
normalized  to  give  rise  to  a  circular  velocity  of  $v_c=220\kms$  at
$R=50\kpc$.}
\end{figure}

\begin{figure}
\ifthenelse{\UseFigs=1}{
\ifthenelse{\CompactFigs=0}
{\includegraphics[width=15cm]{Halo.fig07.ps}}
{\includegraphics[width=15cm]{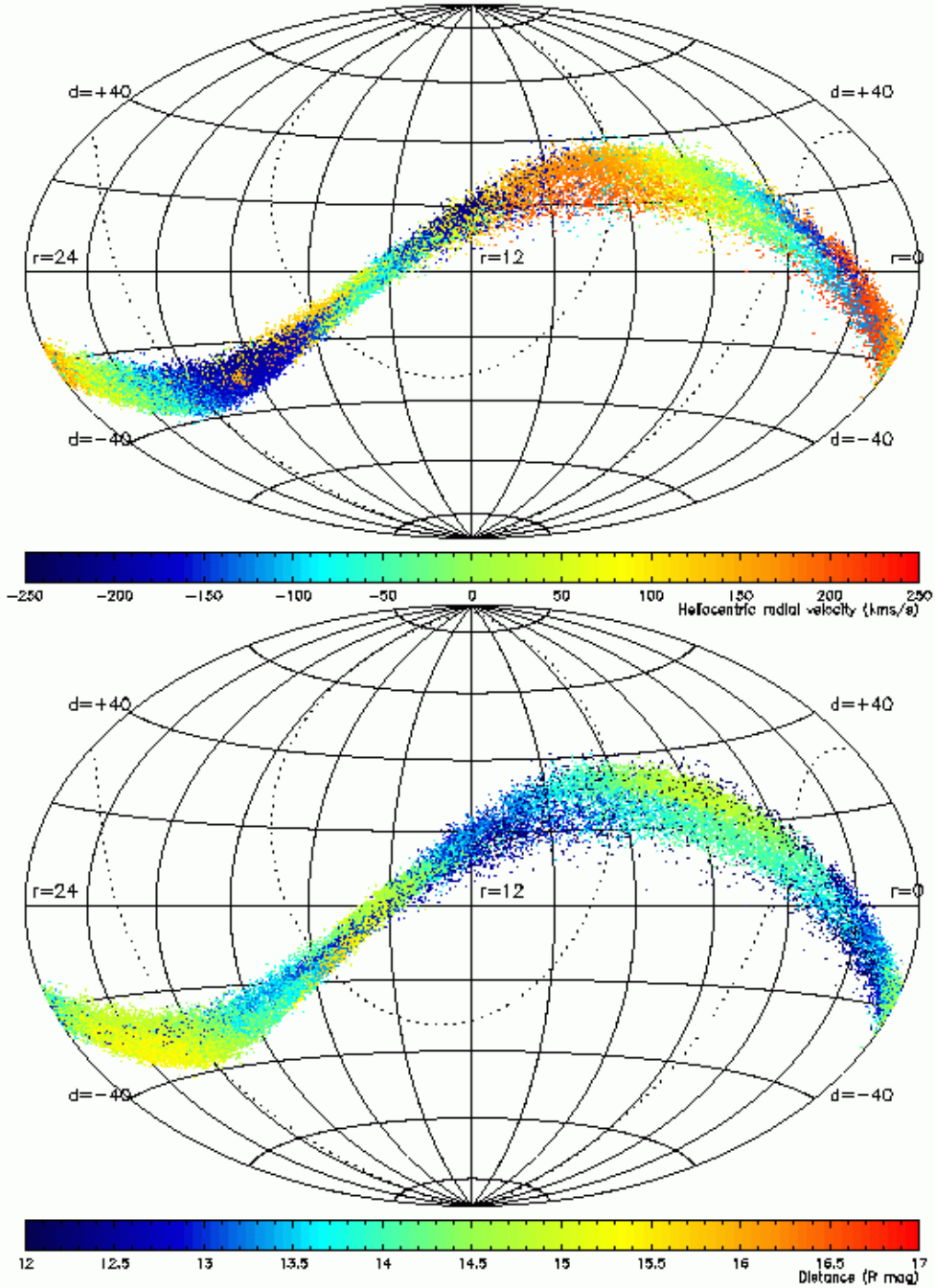}}}{}
\figcaption{The  end-state of the  simulation (at  $T=12\Gyr$) of  the dwarf
galaxy  model D1 (originally  containing $10^5$  particles) in  the Galactic
potential  model  H1  with  $v_c=220\kms$,  and  $q_m=1.0$  is  shown.   The
distribution of debris on the sky follows quite closely a Great Circle.}
\end{figure}

\begin{figure}
\ifthenelse{\UseFigs=1}{
\ifthenelse{\CompactFigs=0}
{\includegraphics[width=15cm]{Halo.fig08.ps}}
{\includegraphics[width=15cm]{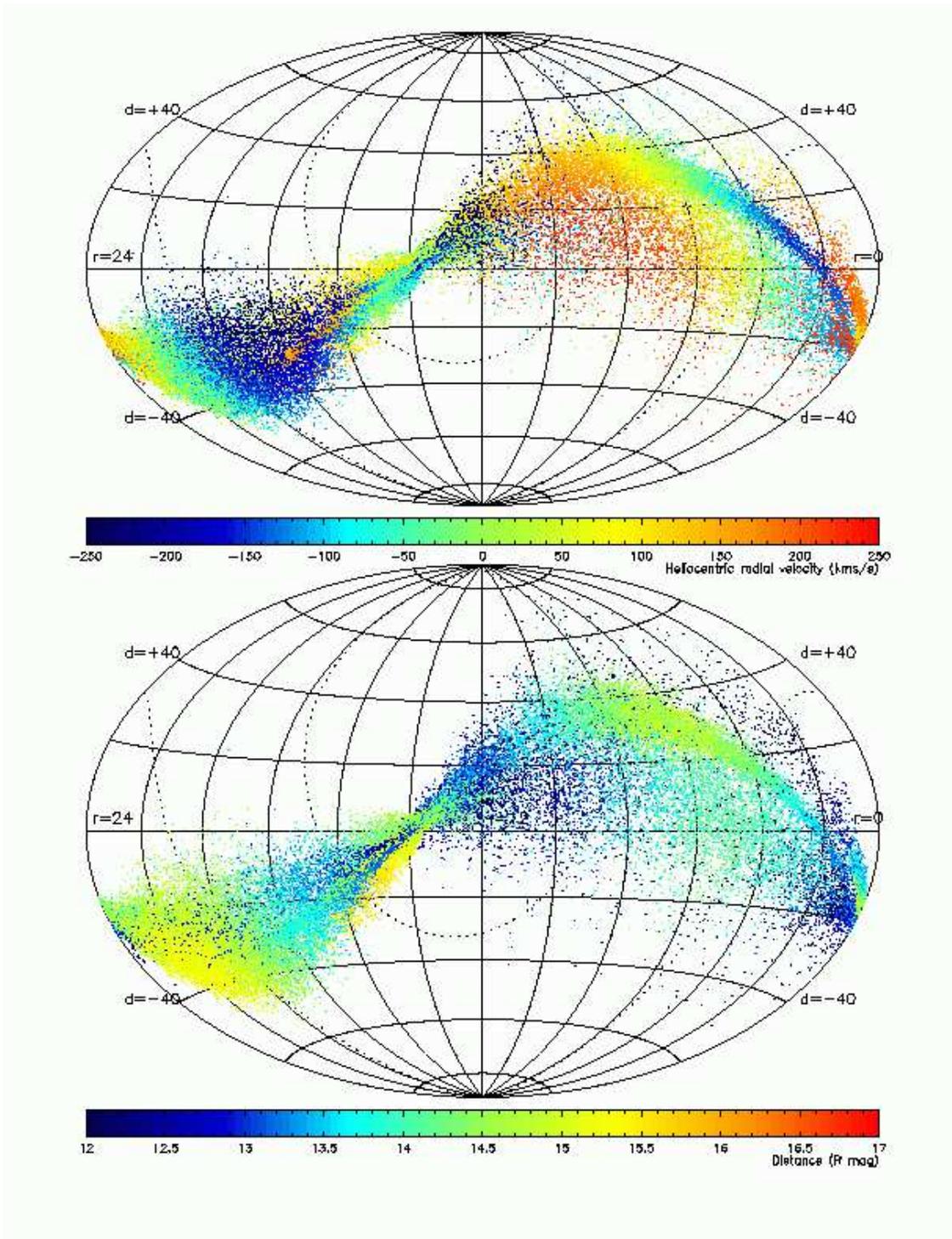}}}{}
\figcaption{Same as  Figure~7, but with  $q_m=0.9$. A significant  amount of
material is now precessed away from the Great Circle stream.}
\end{figure}

\begin{figure}
\ifthenelse{\UseFigs=1}{
\ifthenelse{\CompactFigs=0}
{\includegraphics[width=15cm]{Halo.fig09.ps}}
{\includegraphics[width=15cm]{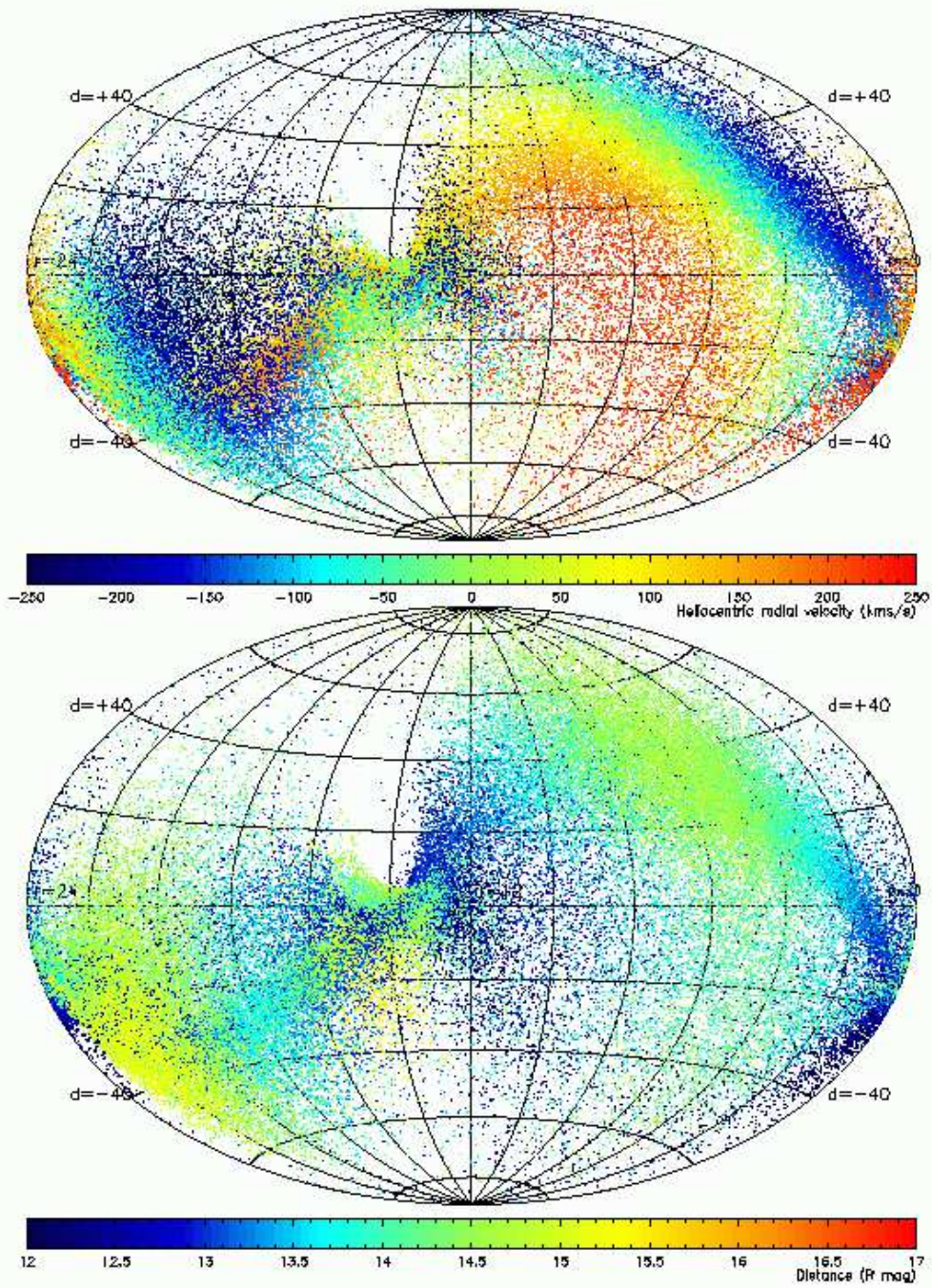}}}{}
\figcaption{Same as Figure~7, but with $q_m=0.75$.}
\end{figure}

\begin{figure}
\ifthenelse{\UseFigs=1}{
\ifthenelse{\CompactFigs=0}
{\includegraphics[width=15cm]{Halo.fig10.ps}}
{\includegraphics[width=15cm]{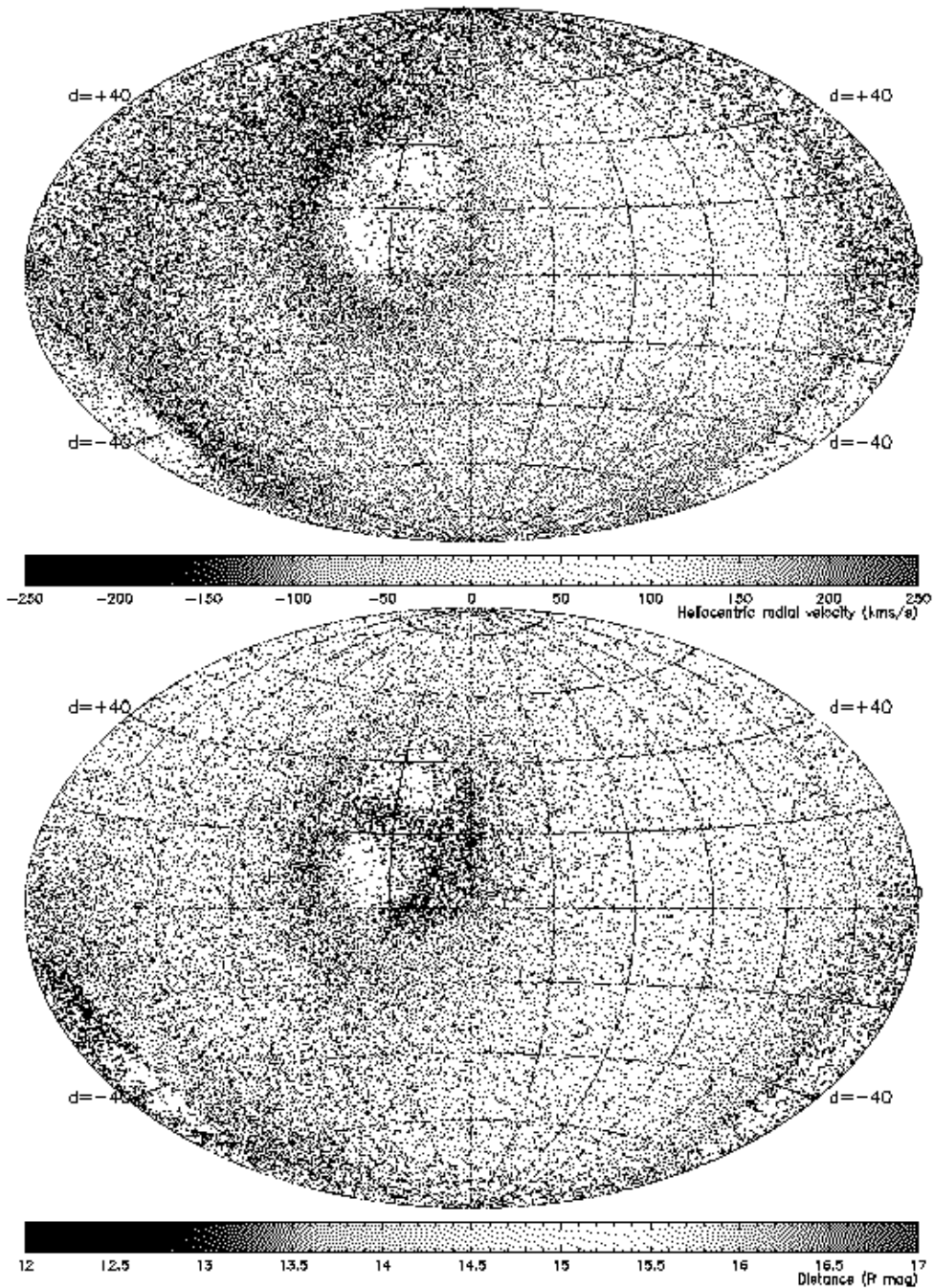}}}{}
\figcaption{Same  as Figure~7,  but with  $q_m=0.5$.  The  tidally disrupted
debris  from  the dwarf  galaxy  has suffered  so  much  precession in  this
flattened potential  that no  sign of a  Great Circle stream  remains. Apart
from the  two obvious holes around  the Galactic poles,  the distribution of
stars on the sky is close  to isotropic. The large velocity dipole, which is
more obvious in this plot than in the preceding three, is simply due to the
Solar reflex motion.}
\end{figure}

\begin{figure}
\ifthenelse{\UseFigs=1}{
\includegraphics[width=12cm]{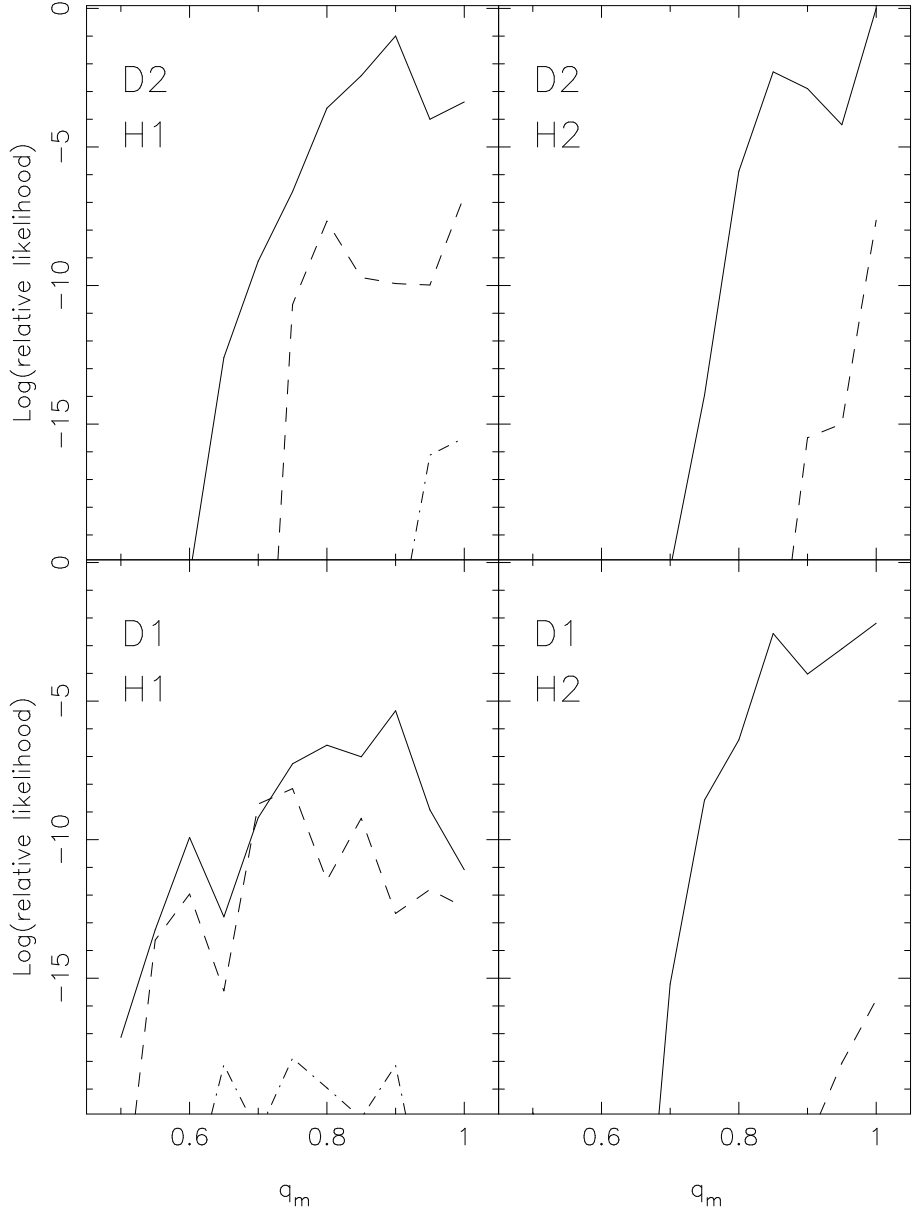}}{}
\figcaption{The relative  likelihoods of the Galactic  models are displayed,
as a function of Halo  mass flattening $q_m$, for the different combinations
of Milky  Way mass model (M1, M2) and  dwarf galaxy model (D1,  D2).  In each
panel, the solid line  corresponds to $v_c(50\kpc)=200\kms$, the dashed line
to $v_c(50\kpc)=220\kms$, and  the dot-dashed line to $v_c(50\kpc)=240\kms$.
The  highest likelihood  model is  the more  massive and  extended  D2 dwarf
galaxy  model simulated  in the  H2 (NFW-type)  Galactic mass  model  with a
spherical Halo.  The  preferred value of the circular  velocity parameter is
always  $v_c(50\kpc)=200\kms$, while  the preferred  shape is  always nearly
spherical.}
\end{figure}

\begin{figure}
\ifthenelse{\UseFigs=1}{
\includegraphics[width=12cm]{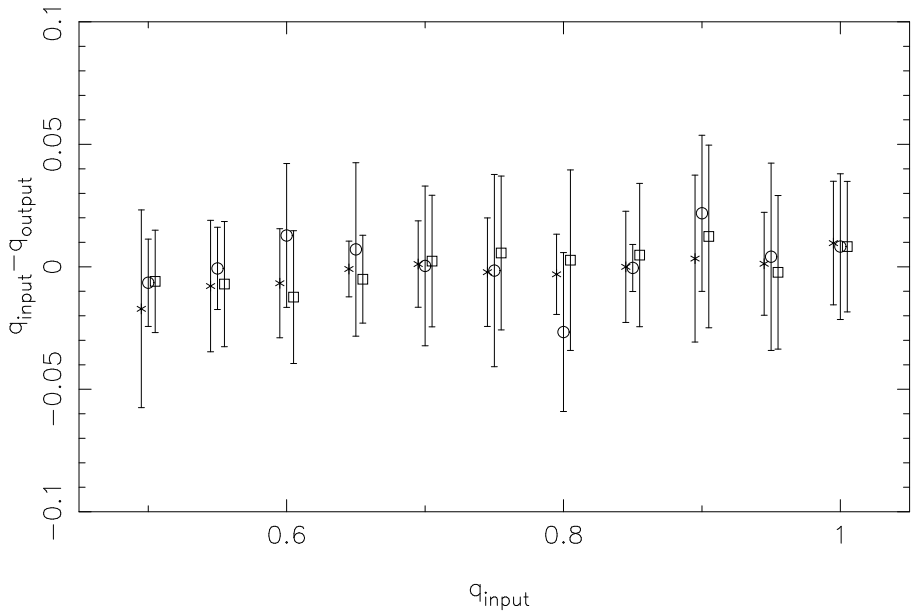}}{}
\figcaption{The difference between input  and recovered Halo mass flattening
$q_m$ as  a function  of input  mass flattening, for  the Monte  Carlo tests
described in \S6.5.1. The cross, circle and square graph marker symbols show
the  mean flattening  difference  for  the simulations  in  which the  input
circular  velocity  was,   respectively,  $v_c=200\kms$,  $v_c=220\kms$  and
$v_c=240\kms$.  (The input  values  of  $q_m$ on  the  abscissa are  shifted
slightly to allow visualization).}
\end{figure}

\begin{figure}
\ifthenelse{\UseFigs=1}{
\includegraphics[width=12cm]{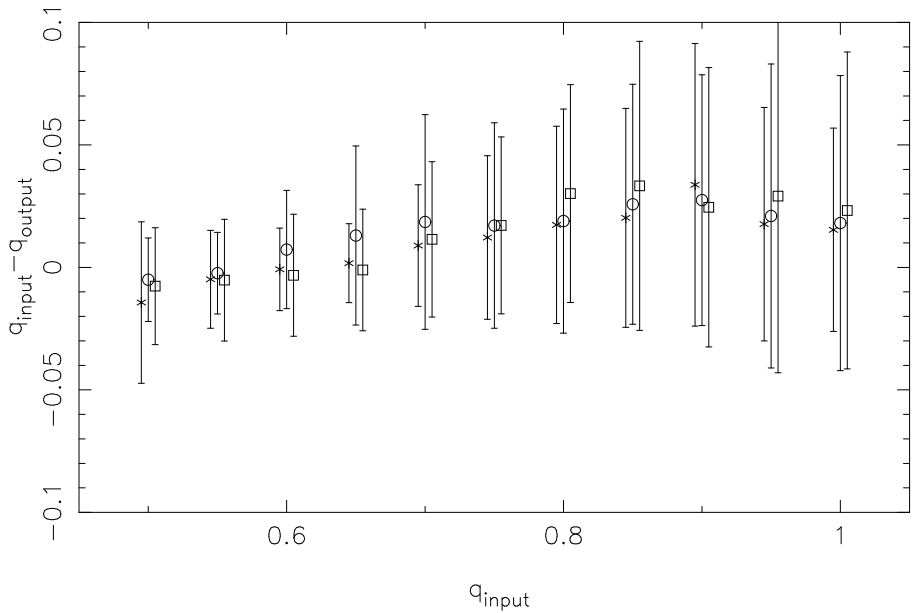}}{}
\figcaption{Same as Figure~12, but for the test described in \S6.5.3.}
\end{figure}

\end{document}